\documentclass[preprint]{aastex}
\newcommand       \be           {\begin{equation}}
\newcommand       \ee           {\end{equation}}
\newcommand       \Angstrom     {\,{\rm \AA}}          

\newcommand       \K            {\,{\rm K}}
\newcommand       \cm           {\,{\rm cm}}

\newcommand	  \g		{\,{\rm g}}

\newcommand	  \yr		{\,{\rm yr}}
\newcommand       \nH           {n_{\rm H}}

\newcommand       \Qext	        {Q_{\rm ext}}

\newcommand       \gtsim        {\gtrsim}
\newcommand       \ltsim        {\lesssim}

\newcommand	  \chithree	{\chi}

\countdef\decade=200
\advance\decade by \year
\countdef\hours=201
\advance\hours by \time
\divide\hours by 60
\countdef\mins=202
\advance\mins by \hours
\multiply\mins by 60
\multiply\hours by 100
\countdef\miltime=203
\advance\miltime by \hours
\advance\miltime by \time
\advance\miltime by -\mins
\def\today{\number\decade.\number\month.\number\day.\number\miltime}
\shorttitle{Grain Size Distributions}
\shortauthors{Weingartner \& Draine}

\begin{document}
\title{Dust Grain Size Distributions and Extinction\\
	in the Milky Way, LMC, and SMC\\
	{\small DRAFT: \today}}

\author{Joseph C. Weingartner}
\affil{Physics Dept., Jadwin Hall, Princeton University,
        Princeton, NJ 08544, USA; CITA, 60 St. George Street, University of
Toronto, Toronto, ON M5S 3H8, Canada; weingart@cita.utoronto.ca}

\and

\author{B.T. Draine}
\affil{Princeton University Observatory, Peyton Hall,
        Princeton, NJ 08544, USA; draine@astro.princeton.edu}

\begin{abstract}

We construct size distributions for carbonaceous and silicate grain 
populations in different regions of the Milky Way, LMC, and SMC.
The size distributions include sufficient very small carbonaceous
grains (including polycyclic aromatic hydrocarbon molecules) to account for
the observed infrared and microwave emission from the diffuse interstellar
medium.  Our distributions reproduce the observed extinction
of starlight, which varies depending upon the 
interstellar environment through which the light travels.  As shown by 
Cardelli, Clayton \& Mathis in 1989, these variations can be roughly 
parameterized by the ratio of visual extinction to reddening, $R_V$.  
We adopt a fairly simple functional form for the size distribution,
characterized by several parameters.  We tabulate these parameters for
various combinations of values for $R_V$ and $b_{\rm C}$, the C abundance 
in very small grains.  We also find size distributions for the line of
sight to HD 210121, and for sightlines in the LMC and SMC.
For several size distributions, we
evaluate the albedo and scattering asymmetry parameter, and present
model extinction curves extending beyond the Lyman limit.  

\end{abstract}


\section{Introduction}

Mathis, Rumpl, \& Nordsieck (1977, MRN) constructed their classic 
interstellar dust model on the basis of the observed extinction of 
starlight for lines of sight passing through diffuse clouds.  Strong 
absorption is observed at $9.7$ and $18 \micron$, corresponding to stretching
and bending modes in silicates.  The strong extinction feature at $2175
\Angstrom$ can be approximately reproduced by small graphite particles
(Stecher \& Donn 1965; Wickramasinghe \& Guillaume 1965).  The simplest 
model incorporating both silicate and graphite material consists of two
separate grain populations, one of silicate composition and one of graphite 
composition.  MRN found that the extinction curve (i.e. the functional 
dependence of the extinction on the wavelength $\lambda$) 
is well reproduced if the grain size distribution 
(with identical form for each component) is given by
\be
dn_{\rm gr} = C \nH a^{-3.5} da,~~~~~~~ a_{\rm min} < a < a_{\rm max}
\label{eq:mrn}
\ee
with $a_{\rm min} = 50 \Angstrom$ and $a_{\rm max} = 0.25 \micron$;
$n_{\rm gr} (a)$ is the number density of grains with size $\le a$ and  
$\nH$ is the number density of H nuclei (in both atoms and molecules).  MRN 
adopted spherical grains, for which Mie theory can be used to compute 
extinction cross sections, and we shall do the same; in this case $a$ is 
the grain radius.  Draine \& Lee (1984) extended the wavelength coverage of
the MRN model, constructed dielectric functions for ``astronomical silicate''
and graphite, and found the following normalizations for the size 
distribution:  
$C = 10^{-25.13}$ ($10^{-25.11}$) $\cm^{2.5}$ 
for graphite (silicate).

Since the development of the MRN model, more observational evidence has 
become available; some of these new observations require revisions of the 
model.  First, the extinction curve has been found to 
vary, depending upon the interstellar environment through which the starlight
passes.  Cardelli, Clayton, \& Mathis (1989, CCM) found that this dependence
can be characterized fairly well by a single parameter,
which they took to be $R_V \equiv A(V)/E(B-V)$, the ratio of visual 
extinction to reddening.  CCM have fitted the average
extinction curve $A(\lambda)/A(V)$ as functions of $\lambda$ and
$R_V$.  For the diffuse ISM, $R_V \approx 3.1$; higher values are observed
for dense clouds.  Kim, Martin, \& Hendry (1994) used the maximum entropy 
method to find smooth size distributions, for silicate and graphite grains,
for which the extinction for $R_V =3.1$ and $5.3$ is well reproduced.  Their
$R_V=5.3$ distribution has significantly fewer ``small'' grains 
($a < 0.1 \micron$) than their $R_V=3.1$ distribution, as well as a 
modest increase at larger sizes.  This result was expected, since generally
there is relatively less extinction at short wavelengths (provided by 
small grains) for larger values of $R_V$.  

Observations of thermal emission from dust have provided another challenge
to the MRN model.  Emission in the 3 to $60 \micron$ range,
presumably generated by grains small enough to reach
temperatures of 30 to $600 \K$ or more upon the absorption of a single
starlight photon (see e.g. Draine \& Anderson 1985), imply a population 
of very small grains (with $a < 50 \Angstrom$).  The non-detection of
the $10 \micron$ silicate feature in emission from diffuse clouds
(Mattila et al. 1996; Onaka et al. 1996) appears to rule out silicate grains
as a major component of the $a \lesssim 15 \Angstrom$ population (but see
note added in proof).  Emission features at 3.3, 6.2, 7.7, 8.6,
and $11.3 \micron$ (see Sellgren 1994 for a review) have been identified 
as C-H and C-C stretching and bending modes in polycyclic aromatic 
hydrocarbons (L\'{e}ger \& Puget 1984), suggesting that the carbonaceous 
grain population extends down into the molecular regime.  
Recent 
observations of dust-correlated microwave emission has been attributed to
the very small grain population (Draine \& Lazarian 1998a).

The abundance of very small grains required to generate the observed IR 
emission from the diffuse ISM is not yet well-known.  In the model of 
D\'{e}sert, Boulanger, \& Puget (1990), polycyclic aromatic hydrocarbon (PAH)
molecules with less than 540 C atoms (equal to the number of C atoms in a 
spherical graphite grain with $a \approx 10 \Angstrom$) lock up a C 
abundance\footnote{By ``abundance'', we mean the number of atoms of an element
per interstellar H nucleus.} 
of $\approx 4 \times 10^{-5}$.  
Li \& Draine (2001) compare
observations of diffuse galactic emission with detailed
model calculations for grains heated by galactic starlight and find
that a C abundance
$\sim 4$--$6 \times 10^{-5}$ is required in hydrocarbon molecules with
$\ltsim10^3$ C atoms.  
They conclude that the emission is best reproduced
if the very small grain population is 
the sum of two log-normal size distributions:\footnote{
	The log-normal distribution with $a_{0,1}=3.5\Angstrom$ is required
	to reproduce the observed 3--25$\micron$ emission, and the
	$a_{0,1}=30\Angstrom$ component is needed to contribute
	emission in the DIRBE $60\micron$ band.
	}
\be
\label{eq:lognormal}
\frac{1}{\nH} \left(\frac{dn_{\rm gr}}{da}\right)_{vsg} 
\equiv D(a) = \sum_{i=1}^2
\frac{B_i}{a} 
\exp \left\{ - \frac{1}{2} \left[ \frac{\ln (a/a_{0,i})}
{\sigma} \right]^2 \right\}
~~~,~~~~~~~a > 3.5 \Angstrom 
\ee
\be
\label{eq:B}
B_i = \frac{3}{(2\pi)^{3/2}}
\frac{\exp(-4.5\sigma^2)}{\rho a_{0,i}^3\sigma}
\frac{
        b_{{\rm C},i}m_{\rm C} 
        }
        {
        1 + {\rm erf}[3 \sigma / \sqrt{2} + 
        \ln (a_{0,i} / 3.5 \Angstrom) / \sigma \sqrt{2}]
        }~~~,
\ee
where
$m_{\rm C}$ is the mass of a C atom,
$\rho=2.24\g\cm^{-3}$ is the density of graphite,
$b_{{\rm C},1}=0.75b_{\rm C}$, $b_{{\rm C},2}=0.25b_{\rm C}$,
$b_{\rm C}$ is the total C abundance (per H nucleus) in the 
log-normal
populations, $a_{0,1} = 3.5 \Angstrom$,
$a_{0,2} = 30 \Angstrom$, and $\sigma = 0.4$.

Draine \& Lazarian (1998b) estimated the electric dipole radiation from 
spinning grains, and found that it could account for the dust-correlated
component of the diffuse Galactic
microwave emission if $b_{\rm C} \approx 2 \times 10^{-5}$.
More recent modelling confirms that the microwave emission can be
reproduced with $b_{\rm C} \approx 2$--$4 \times 10^{-5}$ 
(B.~T.~Draine \& A.~Li 2001, in preparation).

Our goal here is to find size distributions which include very small 
carbonaceous grains\footnote{We take ``carbonaceous grains'' to refer to
graphitic grains and PAH molecules.  Although not all carbonaceous grains
are graphite, we will continue to refer to the dust model considered here
as the ``graphite/silicate'' model, for simplicity.} 
(in numbers sufficient to explain the observed infrared and microwave
emission attributed to this population)
and are consistent with the observed extinction, for
different values of $R_V$ 
in the local Milky Way and for regions in the Large and Small
Magellanic Clouds.
We consider several values of $b_{\rm C}$,
since the C abundance in very small grains is not yet established.
We discuss the observational constraints 
and our method for fitting the extinction in \S 2, present results 
in \S 3, and give a discussion in \S 4.  
The size distributions obtained here will be employed in separate studies,
including an investigation of photoelectric heating by interstellar dust
(Weingartner \& Draine 2001).

\section{Fitting the Extinction}

\subsection{``Observed'' Extinction \label{sec:Aobs}}

For the ``observed'' extinction $A_{\rm obs}(R_V, \lambda)$, 
we adopt the parametrization given by Fitzpatrick (1999).
Bohlin, Savage, \& Drake (1978) found that the ratio of
the total neutral hydrogen column density $N_{\rm H}$
(including both atomic and molecular forms) 
to $E(B-V)$ is fairly constant for the diffuse ISM, with value 
$5.8 \times 10^{21} \cm^{-2}$.  This provides the normalization for the 
extinction curve:  $A(V)/N_{\rm H} = 5.3 \times 10^{-22} \cm^2$.  
The normalization is less clear for dense clouds, because of the difficulty
in measuring $N_{\rm H}$.\footnote{  
Kim \& Martin (1996) compiled a set of sight lines for which both
$A(V)/N_{\rm H}$ and $R_V$ are observationally determined.  Their data are
consistent with $A(V)/N_{\rm H}$ being independent of $R_V$, but the 
uncertainties are large.}
CCM found that $A(\lambda)/A(I)$ 
appears to be independent of $R_V$ for $\lambda >  0.9 \micron$ 
(= {\it I} band), suggesting
that the diffuse cloud value of $A(I)/N_{\rm H} = 2.6 \times 10^{-22}
\cm^2$ may also hold for dense clouds (see, e.g., Draine 1989); we adopt
this normalization. 

\subsection{Functional Form for the Size Distribution}

Lacking a satisfactory theory for the size distribution of interstellar
dust, we employ functional forms for the distribution which
(1) allow for a smooth cutoff for size $a > a_t$, with control of
the steepness of this cutoff; and 
(2) allow for a change in the slope $d\ln n_{\rm gr}/d\ln a$ for $a < a_t$.
We adopt the following form:
\be
\frac{1}{\nH} \frac{dn_{\rm gr}}{da} = D(a) + \frac{C_{\rm g}}{a} 
\left( \frac{a}{a_{\rm t,g}} \right)^{\alpha_{\rm g}}
F(a; \beta_{\rm g}, a_{\rm t,g}) \times
\cases{1~~~, &$3.5 \Angstrom < a < a_{\rm t,g}$
\cr
\exp \left\{ - [(a - a_{\rm t,g})/a_{\rm c,g}]^3 \right\}~~~,
&$a > a_{\rm t,g}$\cr}
\label{eqn:gradist}
\ee
for carbonaceous dust [with $D(a)$ from eq.\ (\ref{eq:lognormal})] and 
\be
\frac{1}{\nH} \frac{dn_{\rm gr}}{da} = \frac{C_{\rm s}}{a}
\left( \frac{a}{a_{\rm t,s}} \right)^{\alpha_{\rm s}} 
F(a; \beta_{\rm s}, a_{\rm t,s}) \times
\cases{1~~~, &$3.5 \Angstrom < a < a_{\rm t,s}$
\cr
\exp \left\{ - [(a - a_{\rm t,s})/a_{\rm c,s}]^3 \right\}~~~,
&$a > a_{\rm t,s}$\cr}
\label{eqn:sildist}
\ee
for silicate dust.  
The term
\be
F(a; \beta, a_{\rm t}) 
\equiv \cases{1+\beta a/a_{\rm t}~~~, &$\beta \ge 0$ \cr
\left(1-\beta a/a_{\rm t} \right)^{-1}~~~, &$\beta < 0$\cr}
\ee
provides curvature. 
The form of the exponential cutoff was suggested by Greenberg (1978).
The structure of the size distribution $D(a)$ for the very small carbonaceous
grains has only a mild effect on the extinction for the wavelengths of 
interest; 
we adopt the same values as Li \& Draine (2001) for 
$a_{0,1} =3.5\Angstrom$,
$a_{0,2} =30\Angstrom$, and $\sigma=0.4$, and the same relative populations
in the two log-normal components
($b_{{\rm C},1}=0.75b_{\rm C}$, $b_{{\rm C},2}=0.25b_{\rm C}$),
but will consider different values of $b_{\rm C}$.
Thus equation (\ref{eqn:gradist}) has a total of six adjustable
parameters ($b_{\rm C}$, $C_{\rm g}$, $a_{\rm t,g}$,
$a_{\rm c,g}$, $\alpha_{\rm g}$, $\beta_{\rm g}$), with another five
parameters ($C_{\rm s}$, $a_{\rm t,s}$,
$a_{\rm c,s}$, $\alpha_{\rm s}$, $\beta_{\rm s}$) in equation 
(\ref{eqn:sildist}) for the silicate size distribution.

\subsection{Calculating the Extinction from the Model}

The extinction at wavelength $\lambda$ is given by 
\be
A(\lambda) = (2.5 \pi \log e) \int d\ln a \frac{dN_{\rm gr}(a)}{da} a^3 
Q_{\rm ext} (a,\lambda)~~~,
\label{eqn:ext}
\ee
where $N_{\rm gr}(a)$ is the column density of grains with size $\le a$ and
$Q_{\rm ext}$ is the extinction efficiency factor, which we evaluate 
(assuming spherical grains) using a Mie theory code derived from BHMIE
(Bohren \& Huffman 1983).

We adopt silicate dielectric functions based on the ``astronomical 
silicate'' functions given by Draine \& Lee (1984) and Laor \& Draine (1993),
but differing in the ultraviolet.
The ``astronomical silicate'' dielectric function
$\epsilon=\epsilon_1+i\epsilon_2$
of Draine \& Lee (1984), based on laboratory measurements of crystalline
olivine in the ultraviolet (Huffman \& Stapp 1973), contains a feature at
$6.5 \micron^{-1}$.  Kim \& Martin (1995) have pointed out that this
feature, which is of crystalline origin, is not present in the observed
interstellar extinction or polarization.  We have therefore excised this
feature from $\epsilon_2$ and ``redistributed'' the oscillator strength over 
frequencies between 8 and $10\micron^{-1}$; we then
recomputed $\epsilon_1$ using the
Kramers-Kronig relation (Draine \& Lee 1984).  (The resulting ``smoothed
astronomical silicate'' dielectric functions are available 
at http://www.astro.princeton.edu/$\sim$draine/.)

For carbonaceous grains, we adopt the description given by Li \& Draine
(2001), in which the smallest grains are PAH molecules, the largest grains
consist of graphite, and grains of intermediate size have optical properties
intermediate between those of PAHs and graphite.  For PAHs, Li \& Draine
estimate absorption cross sections per C atom, for both neutral and ionized
molecules.  
Li \& Draine estimate PAH absorption near 2175$\Angstrom$ by
{\it assuming} that the 2175$\Angstrom$ absorption profile is in
large part due to the PAH population; our adopted PAH absorption cross sections
near 2175$\Angstrom$ therefore agree -- by construction -- 
with the observed 2175$\Angstrom$ profile.
We convert to a size-based description by assuming a C density
$\rho = 2.24 \g \cm^{-3}$, and we assume that 50\% are neutral and 50\% are
ionized (the ionization state affects the absorption by these grains at
$\lambda \gtsim 0.6 \micron$).  We take graphite dielectric functions from
Draine \& Lee (1984) and Laor \& Draine (1993) and adopt the usual 
``$1/3-2/3$'' approximation:  $\Qext = [\Qext 
({\epsilon}_{\parallel}) + 2 \Qext ({\epsilon}_{\perp})]/3$,
where ${\epsilon}_{\parallel}$ and 
${\epsilon}_{\perp}$ are the components of the graphite dielectric tensor 
for the electric field 
parallel and perpendicular to the $c$-axis, respectively.  
Draine \& Malhotra (1993) showed that
the $1/3-2/3$ approximation is sufficiently accurate for extinction curve
modelling.  

\subsection{Abundance/Depletion Constraints \label{sec:abun_dep}}

Given estimates of the abundances and interstellar 
depletions of the elements incorporated in dust and the mass densities 
of the grain materials, we can estimate the total volume per H atom, 
$V_{\rm tot}$, in the carbonaceous and silicate grain populations.
For a long time, solar abundances were used for this purpose (see Grevesse
\& Sauval 1998 for a recent compendium of solar abundances).  Recent evidence,
e.g.~from measurements of abundances in the atmospheres of B stars, suggest 
that the abundances in the present-day ISM may be substantially
lower than the solar values 
(see Snow \& Witt 1996; Mathis 1996, 2000; and  Snow 2000 for reviews).  
However, Fitzpatrick \& Spitzer (1996) concluded that S 
has solar abundance in the ISM, and Howk, Savage, \& 
Fabian (1999) found solar abundances of Zn, P, and S along the line of sight
to $\mu$ Columbae.  Thus, interstellar abundances are not yet well-known.  

We adopt the solar C abundance of $3.3 \times 10^{-4}$ (Grevesse \& Sauval
1998) and assume that $\approx 30 \%$ is in the gas phase.\footnote{Cardelli
et al.~(1996) and Sofia et al.~(1997) found a gas-phase C abundance of 
$1.4 \times 10^{-4}$, larger than the $\approx 1 \times 10^{-4}$ that we
assume.}  With the ideal graphite density of $2.24 \g \cm^{-3}$, we find
$V_{\rm tot,g} \approx 2.07 \times 10^{-27} \cm^3 \, {\rm H}^{-1}$ for
carbonaceous dust.  To estimate the total volume in amorphous
silicates, we assume a stoichiometry 
approximating MgFeSiO$_4$, with mass number per structural unit of 172.  
Since Si, Mg, and Fe have similar abundances in the Sun and are all highly
depleted in the ISM (Savage \& Sembach 1996), we simply assume that the Si 
abundance in silicate dust is equal to its solar value of 
$3.63 \times 10^{-5}$.  We adopt a density of $3.5 \g \cm^{-3}$, intermediate
between the values for 
crystalline forsterite (Mg$_2$SiO$_4$, $3.21\g\cm^{-3}$) and 
fayalite (Fe$_2$SiO$_4$, $4.39\g\cm^{-3}$).  
Thus, we estimate
$V_{\rm tot,s} \approx 2.98 \times 10^{-27} \cm^3 \, {\rm H}^{-1}$ for 
silicate dust.

\subsection{Method of Solution \label{sec:soln}}

For a given pair of values ($R_V$, $b_{\rm C}$), 
we seek the best fit to 
the extinction by varying the powers $\alpha_{\rm g}$ and $\alpha_{\rm s}$;
the ``curvature'' parameters $\beta_{\rm g}$ and $\beta_{\rm s}$; 
the transition sizes $a_{\rm t,g}$ and $a_{\rm t,s}$; 
the upper cutoff parameters $a_{\rm c,g}$ and  
$a_{\rm c,s}$; and the total volume per H in both
the carbonaceous
and silicate distributions, $V_{\rm tot,g}$ and $V_{\rm tot,s}$,
respectively.

We use the Levenberg-Marquardt method, as implemented in Press et al.
(1992), to fit the continuous extinction between $0.35 \micron^{-1}$
and $8 \micron^{-1}$.\footnote{The lower limit of $0.35 \micron^{-1}$
	was chosen so as to avoid infrared absorption features, 
	most notably the
	$3.4 \micron$ C-H stretch
	feature.  
	Extinction data for $\lambda^{-1} > 8
	\micron^{-1}$ are very limited.}  
We evaluate the extinction at 
100 wavelengths $\lambda_i$, equally spaced in $\ln \lambda$, 
and minimize one of two error functions.  In the first case (hereafter 
``case A'') we minimize $\chithree^2 = \chi_1^2 + \chi_V^2$.  

The first term in 
$\chithree^2$ 
gives the error in the extinction fit:
\be
\chi_1^2 = \sum_i \frac{\left( \ln A_{\rm obs} - \ln A_{\rm mod} \right)^2}
{\sigma_i^2}~~~,
\ee
where $A_{\rm obs}(\lambda_i)$ is the average ``observed'' extinction
(\S \ref{sec:Aobs}), $A_{\rm mod}(\lambda_i)$ 
is the extinction computed for the model
[equation (\ref{eqn:ext})], and the $\sigma_i$ are weights.  When evaluating
$A_{\rm mod}$, we verify that the integral in equation (\ref{eqn:ext}) is
evaluated accurately.
We take the weights $\sigma_i^{-1} = 1$ for $1.1 \micron^{-1} < 
\lambda^{-1} < 8 \micron^{-1}$ and $\sigma_i^{-1} = 1/3$ for 
$\lambda^{-1} < 1.1 \micron^{-1}$, 
since the actual IR extinction is uncertain.

The term $\chi_V^2$ is a penalty which keeps the total volumes in the 
carbonaceous and silicate grain populations from grossly exceeding the 
abundance/depletion-limited values found in \S \ref{sec:abun_dep}.  We 
take 
\begin{equation}
\chi_V^2 = 0.4 [\max(\tilde{V}_{\rm g},1) -1]^{1.5} + 
0.4 [\max(\tilde{V}_{\rm s},1) -1]^{1.5}
\label{eq:chi_V}~~~,
\end{equation}
where $\tilde{V}_{\rm g} = V_{\rm tot,g}/
2.07 \times 10^{-27} \cm^3 \, {\rm H}^{-1}$ and
$\tilde{V}_{\rm s} = V_{\rm tot,s}/ 2.98 \times 10^{-27} \cm^3 \, 
{\rm H}^{-1}$.  

Given our assumption that $A(I)/N_{\rm H}$ is independent of $R_V$, the 
extinction for higher $R_V$ can be fit using less total grain volume.  
It seems highly unlikely that
material is transferred from grains to the gas phase as gas and dust cycles
into regions of higher density.
Thus, we also consider a second case
(``case B'') for which the grain volumes are held fixed at approximately
the values found for $R_V=3.1$:  $V_{\rm tot,s}=3.9 \times 10^{-27} \cm^{3} \, 
{\rm H}^{-1}$ and $V_{\rm tot,g}=2.3 \times 10^{-27} \cm^{3} \, {\rm H}^{-1}$.
In this case, we seek to minimize $\chi_1^2$.  

\section{Results \label{sec:results}}

\subsection{Dust in the Milky Way}

\subsubsection{Size Distributions and Extinction Fits}

\begin{figure}
\epsscale{1.00}
\plotone{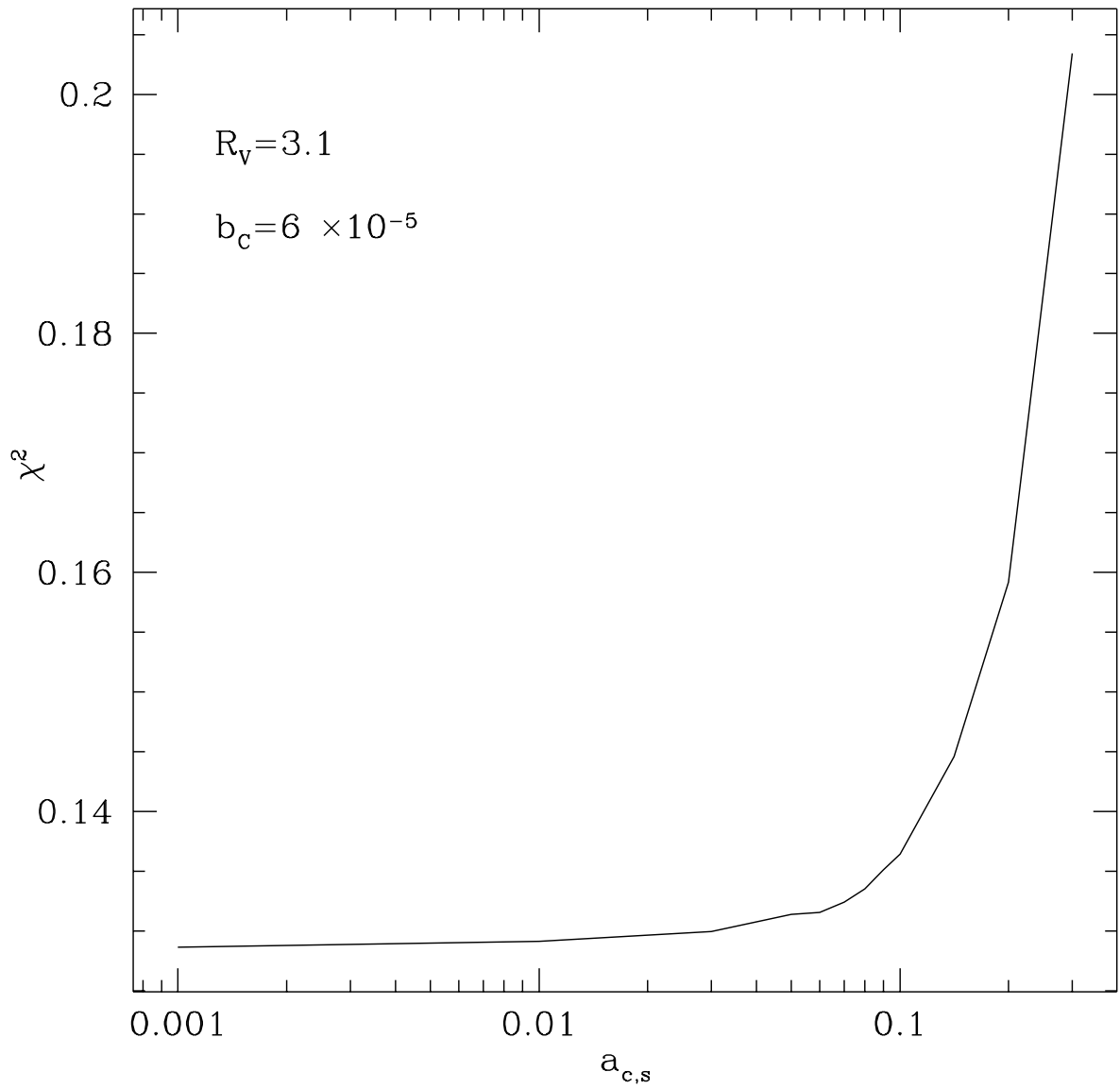}
\caption{
\label{fig:chisq_acs}
The error function $\chi^2$ versus the silicate cutoff parameter,
$a_{\rm c,s}$.
	}
\end{figure}

We have generally found, in fitting the extinction, that $\chi^2$ varies 
only slightly with the silicate cutoff parameter
$a_{\rm c,s}$ until $a_{\rm c,s}$ exceeds a critical 
value of $\approx 0.1 \micron$ (for Milky Way dust; see Figure 
\ref{fig:chisq_acs}).  As $a_{\rm c,s}$ increases, the silicate grains
contribute less short-wavelength extinction, and a large abundance of 
small carbonaceous grains is required to pick up the slack.  When
$a_{\rm c,s} \gtsim 0.1 \micron$, the $2175 \Angstrom$ hump is 
overproduced.  Although $\chi^2$ is nearly constant for 
$a_{\rm c,s} \ltsim 0.1 \micron$, it does increase slightly with 
$a_{\rm c,s}$.  Consequently, our fitting algorithm returns very small
values for $a_{\rm c,s}$, for which the silicate size distribution drops
off very sharply at the large-size end.  Since such sharp cutoffs are
unlikely to occur in nature, we have opted to fix 
$a_{\rm c,s}=0.1 \micron$.  

\begin{figure}
\epsscale{1.00}
\plotone{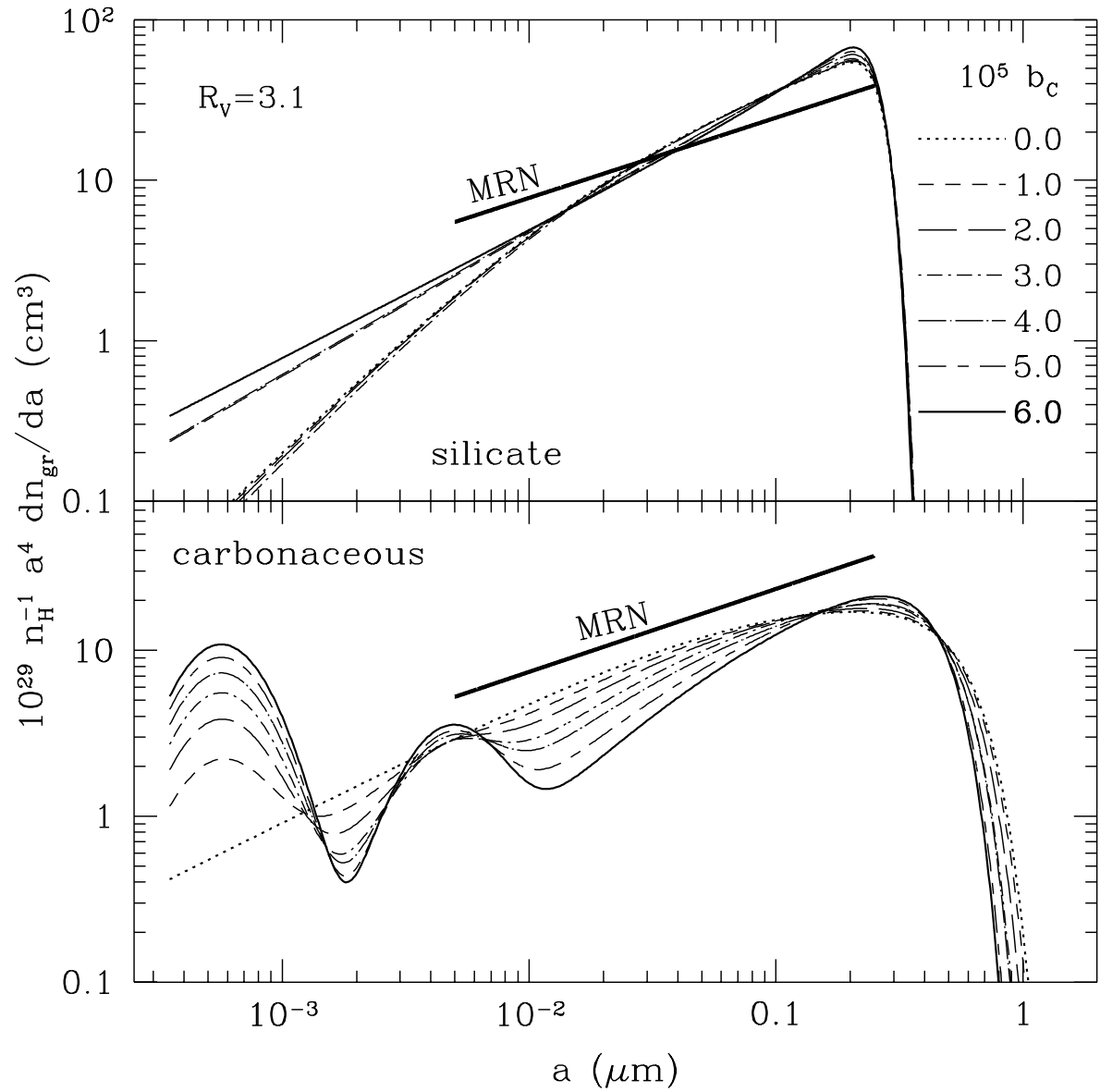}
\caption{
\label{fig:grdist_3.1}
Case A grain size distributions for $R_V=3.1$.  The values of $b_{\rm C}$ are 
indicated.  The heavy, solid lines are the MRN distribution, for comparison.
Our favored distribution has $b_{\rm C}=6\times10^{-5}$ (see text).
	}
\end{figure}

\begin{figure}
\epsscale{1.00}
\plotone{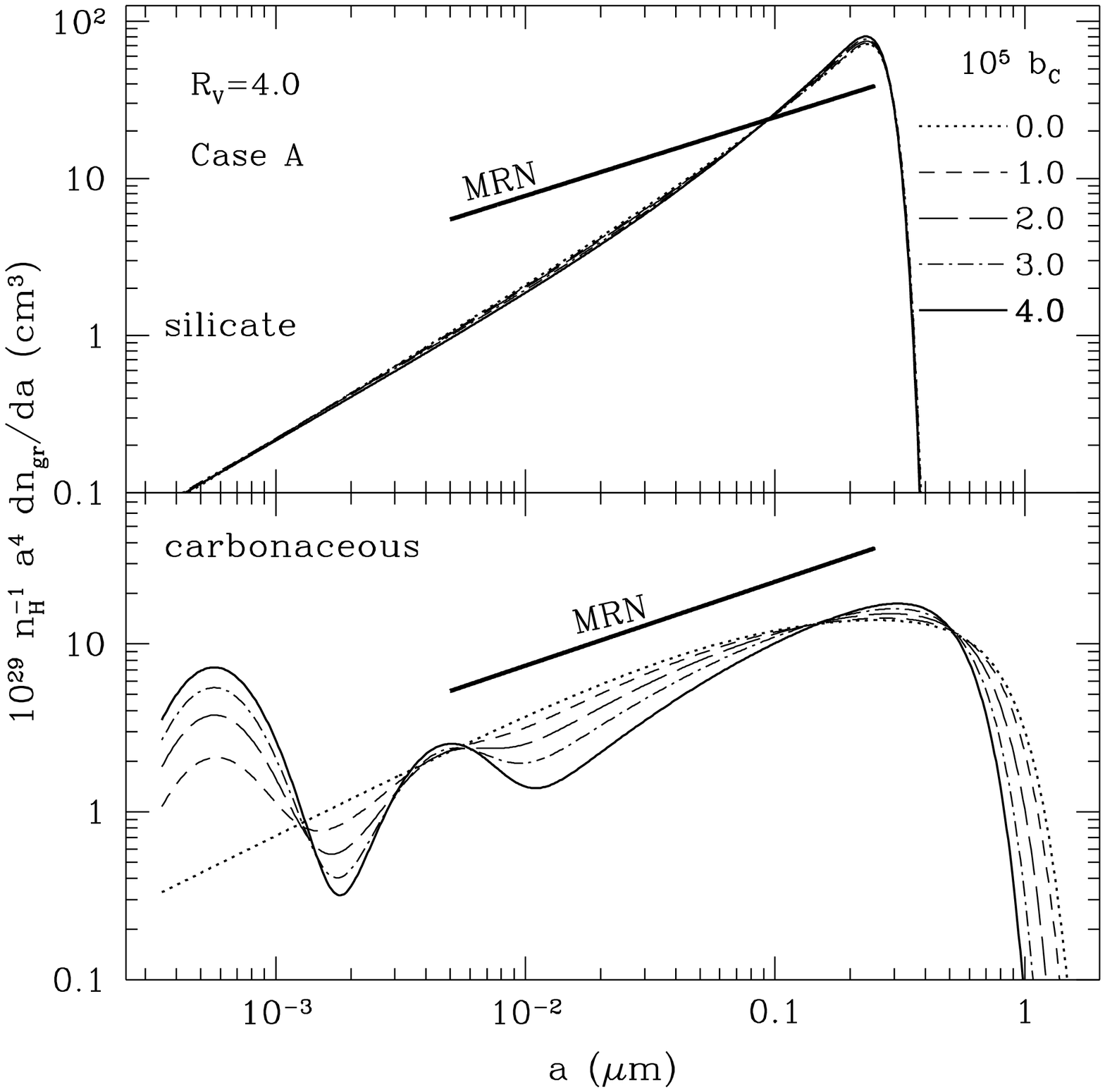}
\caption{
\label{fig:grdist_4.0}
Same as Figure \ref{fig:grdist_3.1}, but for $R_V=4.0$.
Our favored distribution has $b_{\rm C}=4\times10^{-5}$ (see text).
        }
\end{figure}

\begin{figure}
\epsscale{1.00}
\plotone{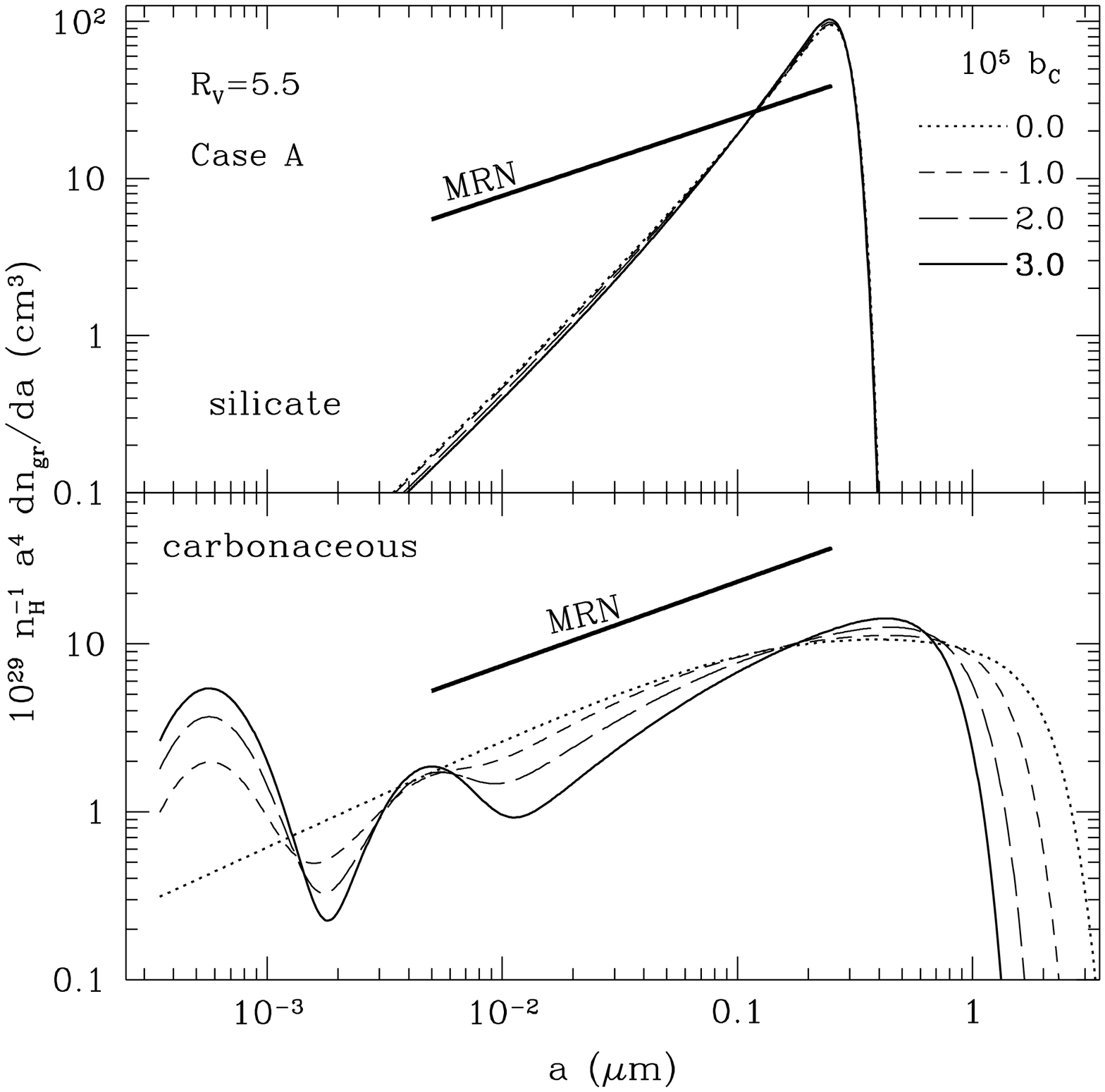}
\caption{
\label{fig:grdist_5.5}
Same as Figure \ref{fig:grdist_3.1}, but for $R_V=5.5$.
Our favored distribution has $b_{\rm C}=3\times10^{-5}$ (see text).
        }
\end{figure}

\begin{figure}
\epsscale{1.00}
\plotone{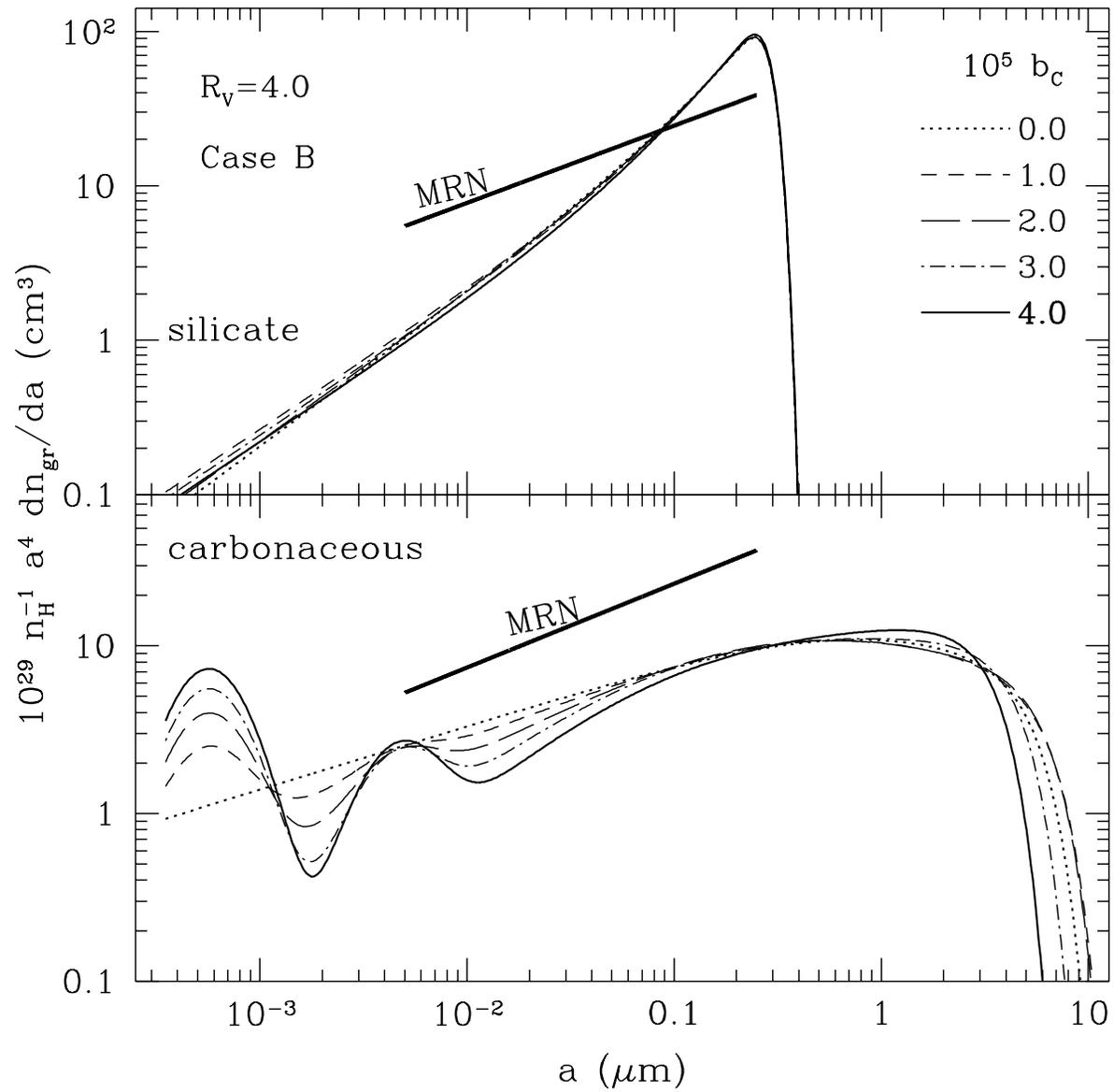}
\caption{
\label{fig:grdist_4.0b}
Case B size distributions for $R_V=4.0$. 
        }
\end{figure}

\begin{figure}
\epsscale{1.00}
\plotone{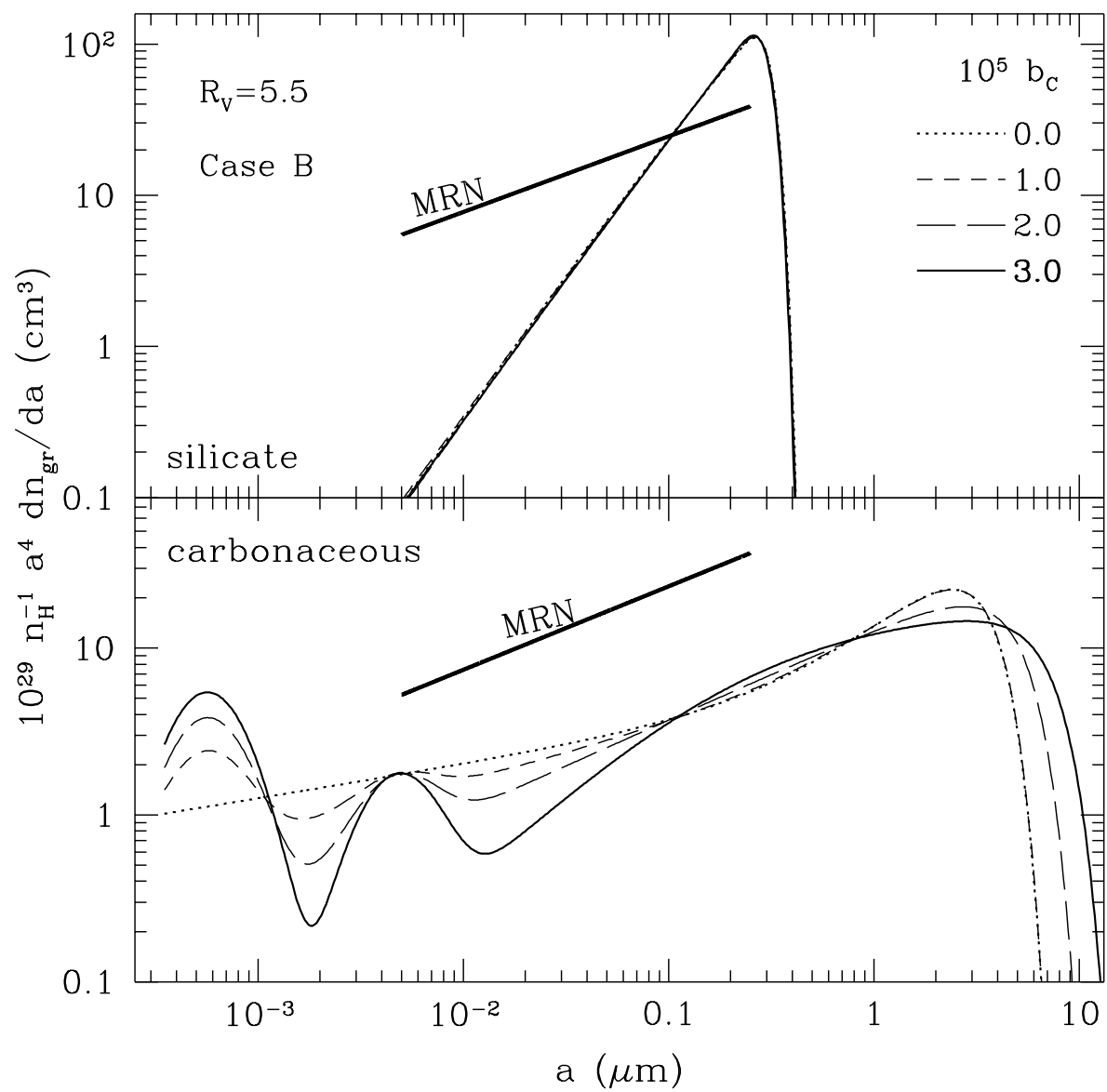}
\caption{
\label{fig:grdist_5.5b}
Case B size distributions for $R_V=5.5$. 
        }
\end{figure}

In Table \ref{tab:grdistpars} we list the values of the distribution 
parameters for which the extinction with $R_V = 3.1$, $4.0$, and $5.5$ 
is best fit, for various values of $b_{\rm C}$.\footnote{These 
parameters [and a FORTRAN subroutine that returns $dn_{\rm gr}/da(a)$] 
are also available in electronic form on the World Wide Web
at www.cita.utoronto.ca/$\sim$weingart.}  
These distributions are 
displayed in Figures \ref{fig:grdist_3.1} through \ref{fig:grdist_5.5b}.

\begin{figure}
\epsscale{1.00}
\plotone{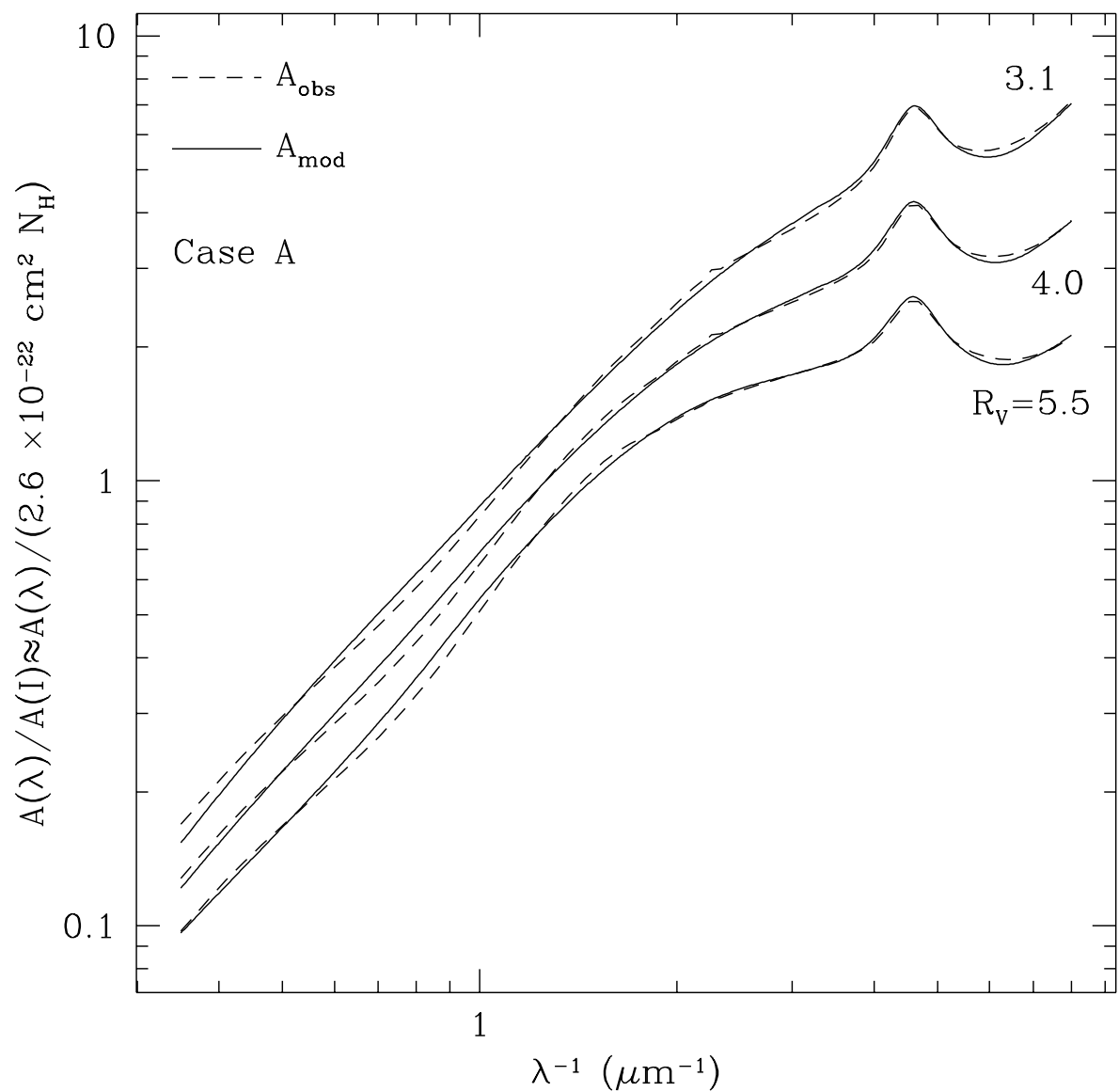}
\caption{
\label{fig:ext_all}
The average ``observed'' extinction $A_{\rm obs}$ and the extinction 
resulting from our case A models for ($R_V$, $10^5 b_{\rm C}$) = (3.1, 6.0),  
(4.0, 4.0), and (5.5, 3.0).  The curves for 
$R_V = 4.0$ ($5.5$) are scaled down by a factor $10^{0.1}$ 
($10^{0.2}$), for clarity.
        }
\end{figure}

\begin{figure}
\epsscale{1.00}
\plotone{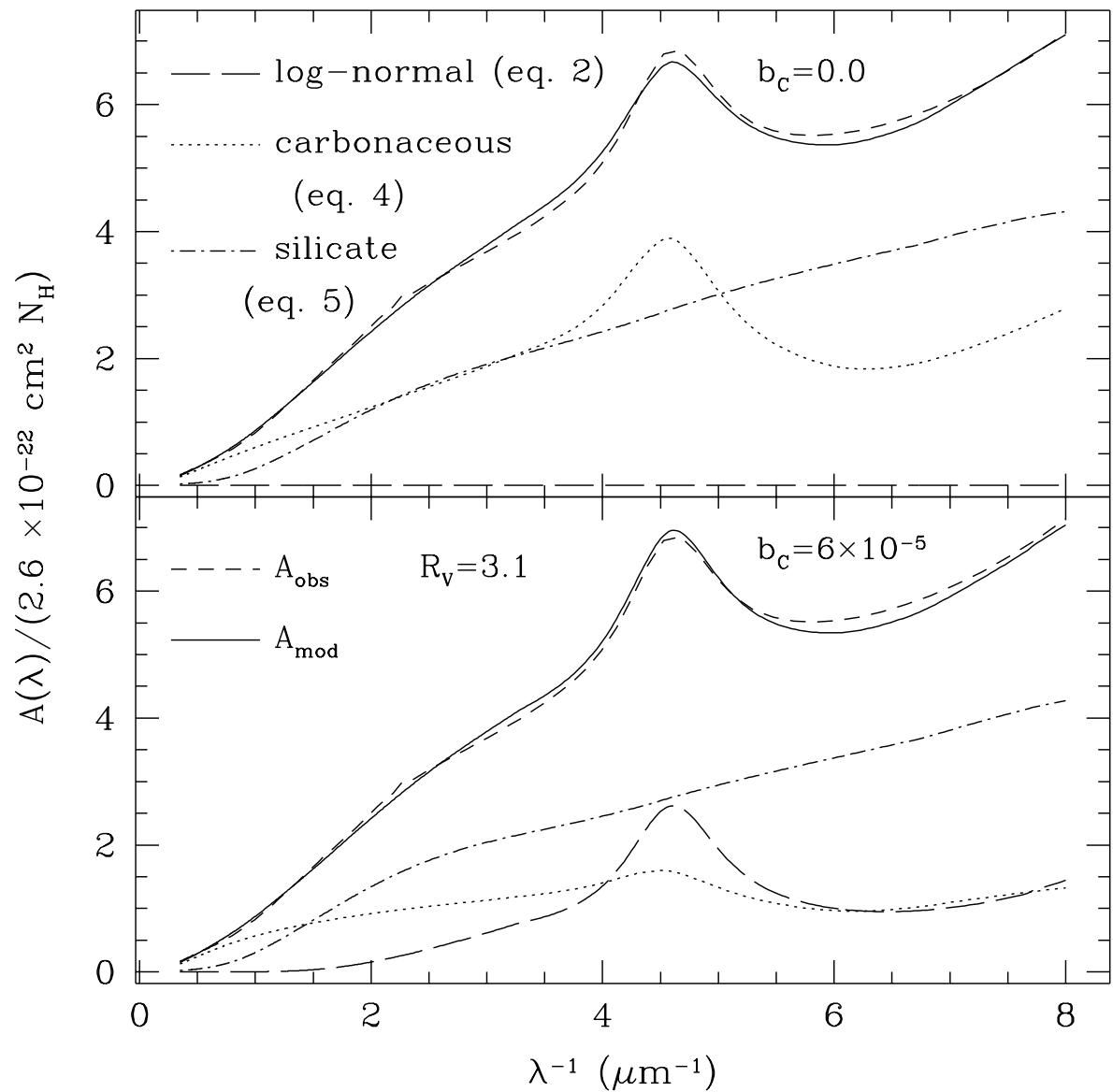}
\caption{
\label{fig:ext_3.1}
The extinction curve $A_{\rm mod}$ resulting from 
the grain distribution of equations 
(\ref{eqn:gradist}) and (\ref{eqn:sildist}), with parameters optimized to fit
$A_{\rm obs}$ (see text) for $R_V = 3.1$ (also shown), for 
$b_{\rm C} = 0.0$ and $6.0 \times 10^{-5}$.  
The contributions from the three grain distribution 
components are also shown.
        }
\end{figure}

\begin{figure}
\epsscale{1.00}
\plotone{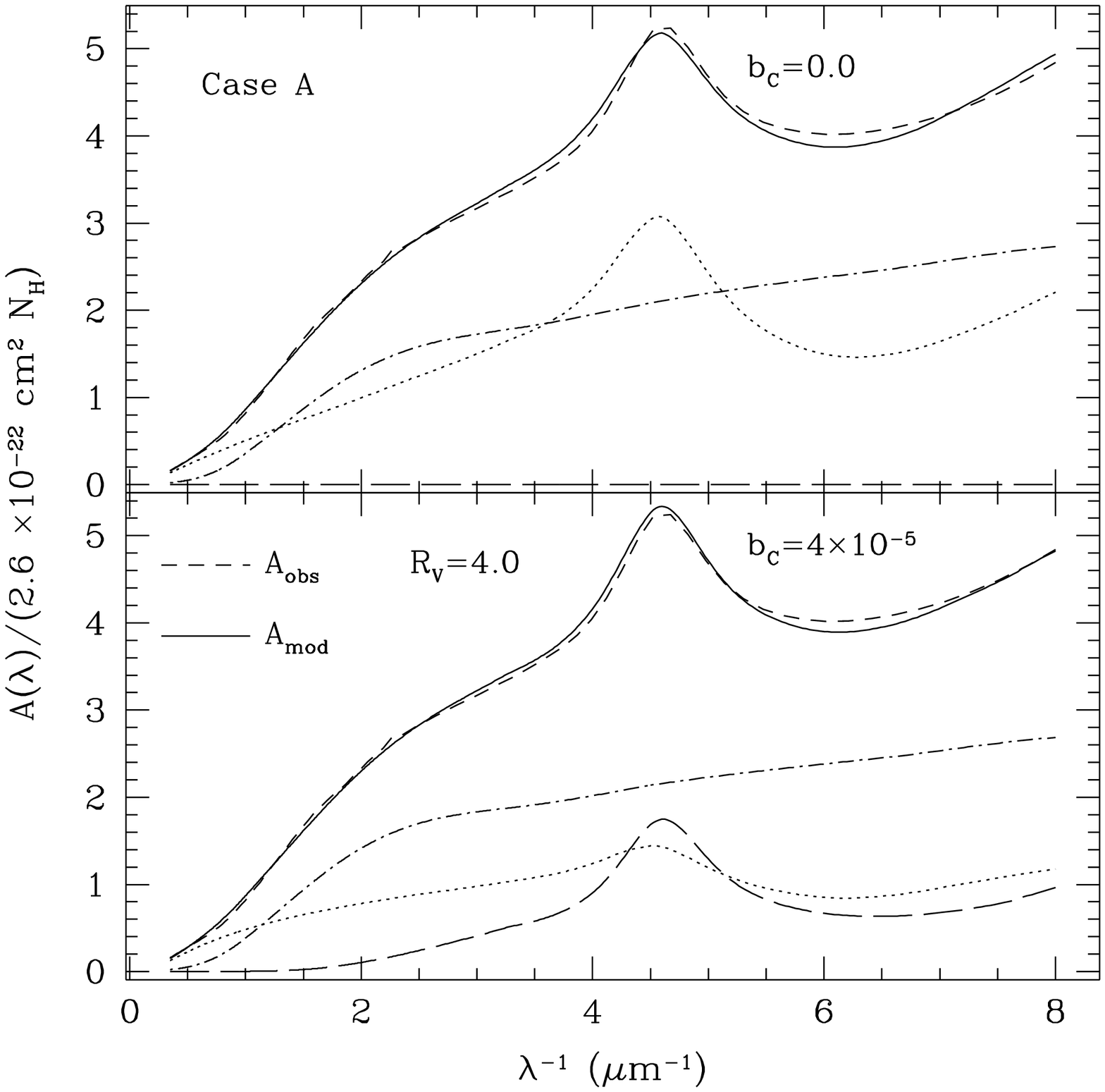}
\caption{
\label{fig:ext_4.0}
Same as Figure \ref{fig:ext_3.1}, but for $R_V = 4.0$ and $b_{\rm C} =
0.0$ and $4.0 \times 10^{-5}$.
        }
\end{figure}

\begin{figure}
\epsscale{1.00}
\plotone{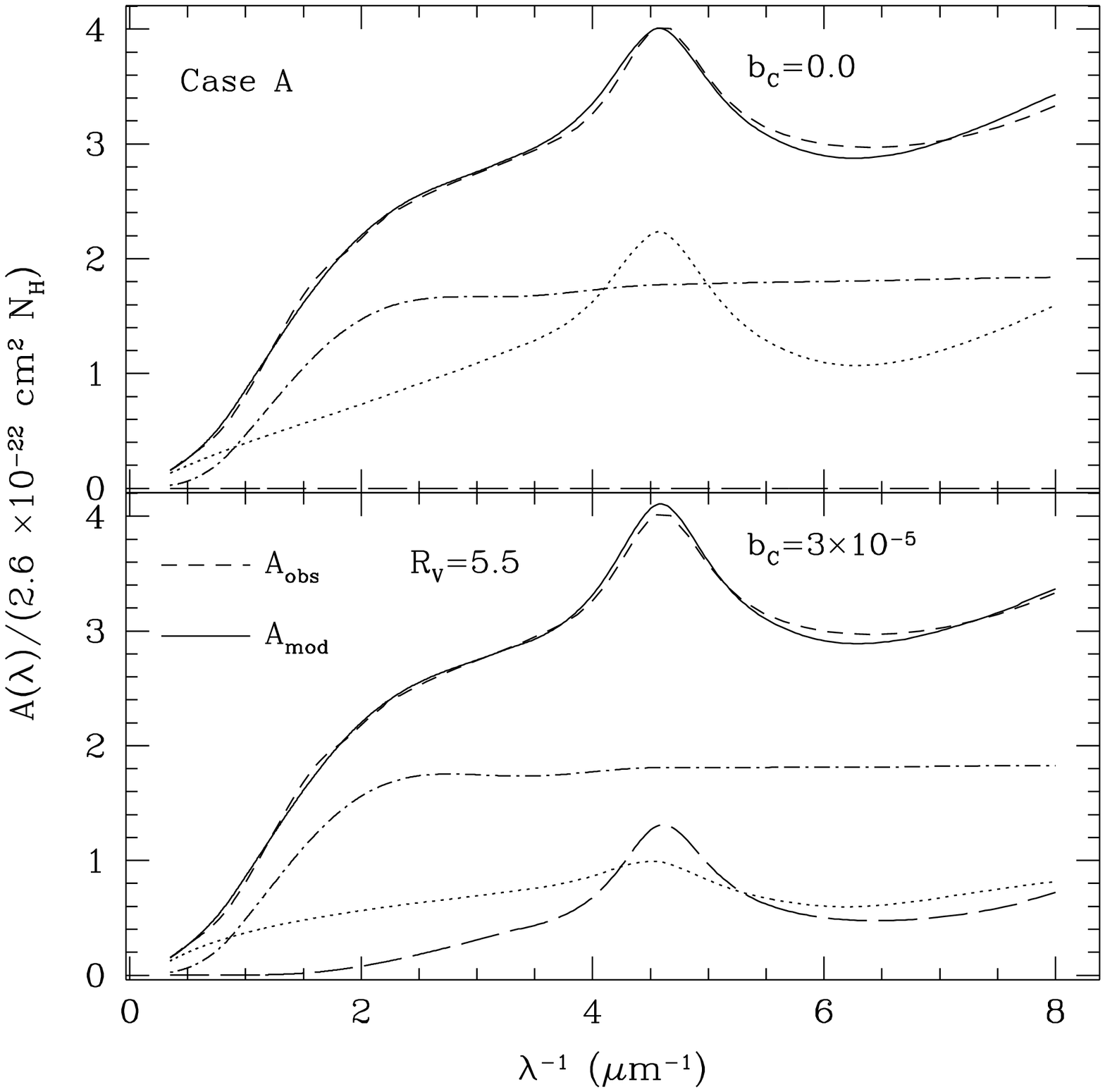}
\caption{
\label{fig:ext_5.5}
Same as Figure \ref{fig:ext_3.1}, but for $R_V = 5.5$ and $b_{\rm C} =
0.0$ and $3.0 \times 10^{-5}$.
        }
\end{figure}

\begin{figure}
\epsscale{1.00}
\plotone{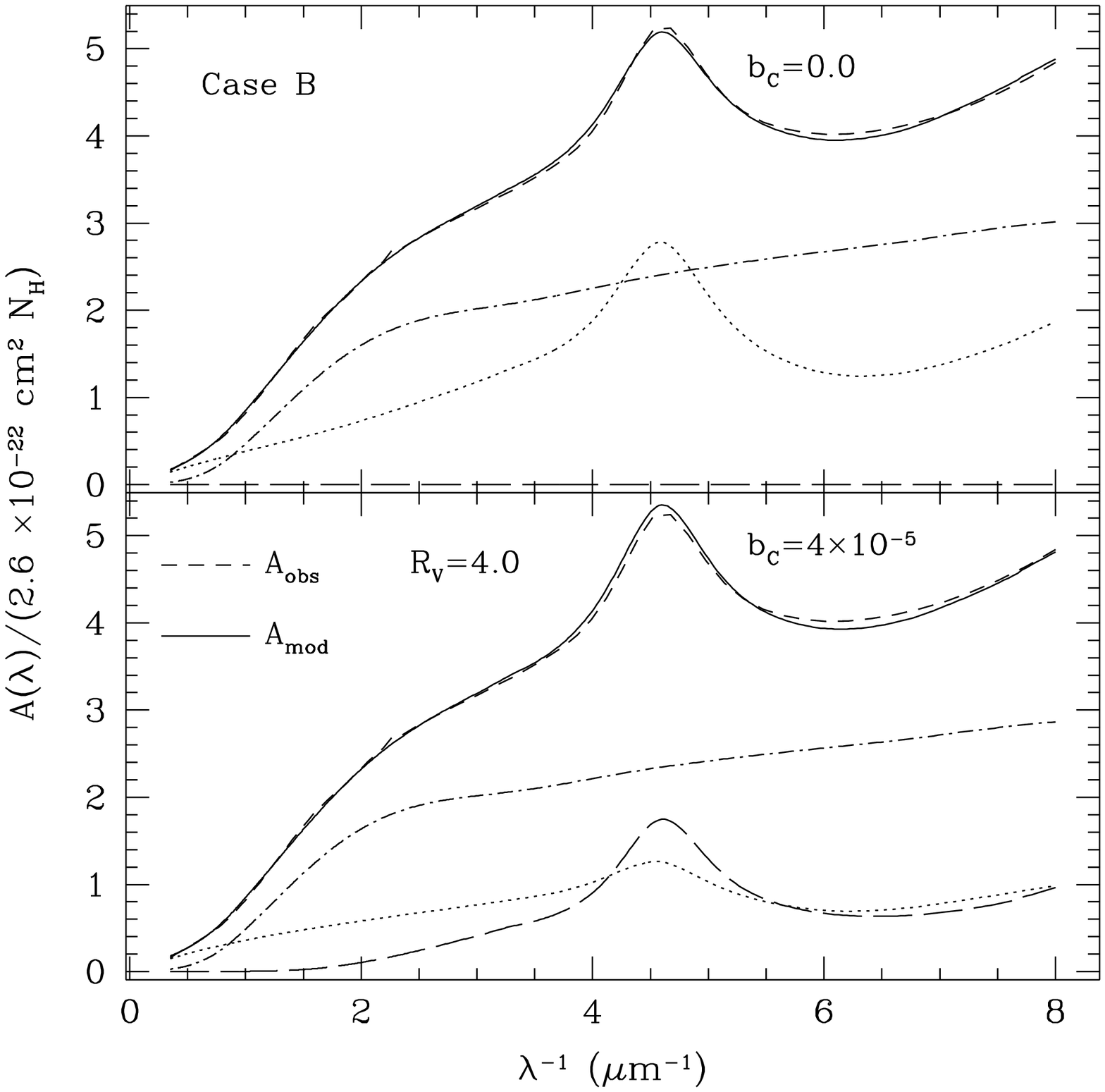}
\caption{
\label{fig:ext_4.0b}
Same as Figure \ref{fig:ext_3.1}, but for $R_V = 4.0$, $b_{\rm C} =
0.0$ and $4.0 \times 10^{-5}$, and fixed total grain volumes
$V_{\rm tot, g}=2.3 \times 10^{-27} \cm^3 \, {\rm H}^{-1}$ and
$V_{\rm tot, s}=3.9 \times 10^{-27} \cm^3 \, {\rm H}^{-1}$.
        }
\end{figure}

\begin{figure}
\epsscale{1.00}
\plotone{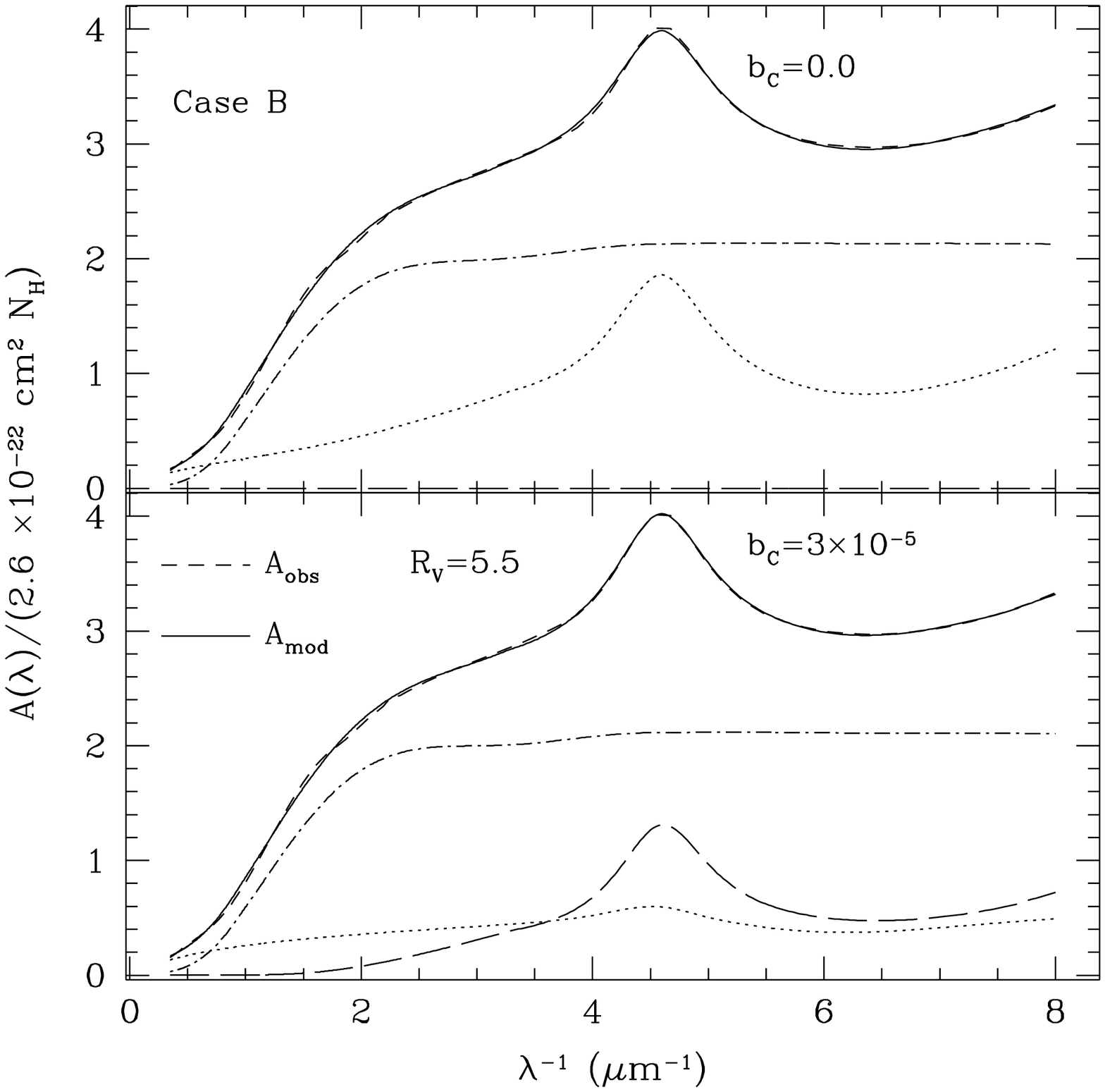}
\caption{
\label{fig:ext_5.5b}
Same as Figure \ref{fig:ext_3.1}, but for $R_V = 5.5$, $b_{\rm C} =
0.0$ and $3.0 \times 10^{-5}$, and fixed total grain volumes
$V_{\rm tot, g}=2.3 \times 10^{-27} \cm^3 \, {\rm H}^{-1}$ and
$V_{\rm tot, s}=3.9 \times 10^{-27} \cm^3 \, {\rm H}^{-1}$.
        }
\end{figure}

In Figure \ref{fig:ext_all} we display $A_{\rm obs}$ and $A_{\rm mod}$ for 
case A, the three values of $R_V$, and the highest values of $b_{\rm C}$ 
included in Table \ref{tab:grdistpars}, 
in a log-log plot, to give a sense for the fit quality
over the entire range of $\lambda^{-1}$.  In Figures \ref{fig:ext_3.1}
through \ref{fig:ext_5.5b}, we display extinction curves
for $b_{\rm C} = 0$ and for the highest value of $b_{\rm C}$ included in
Table \ref{tab:grdistpars}; we show the contribution from each of the
grain distribution components.

\begin{figure}
\epsscale{1.00}
\plotone{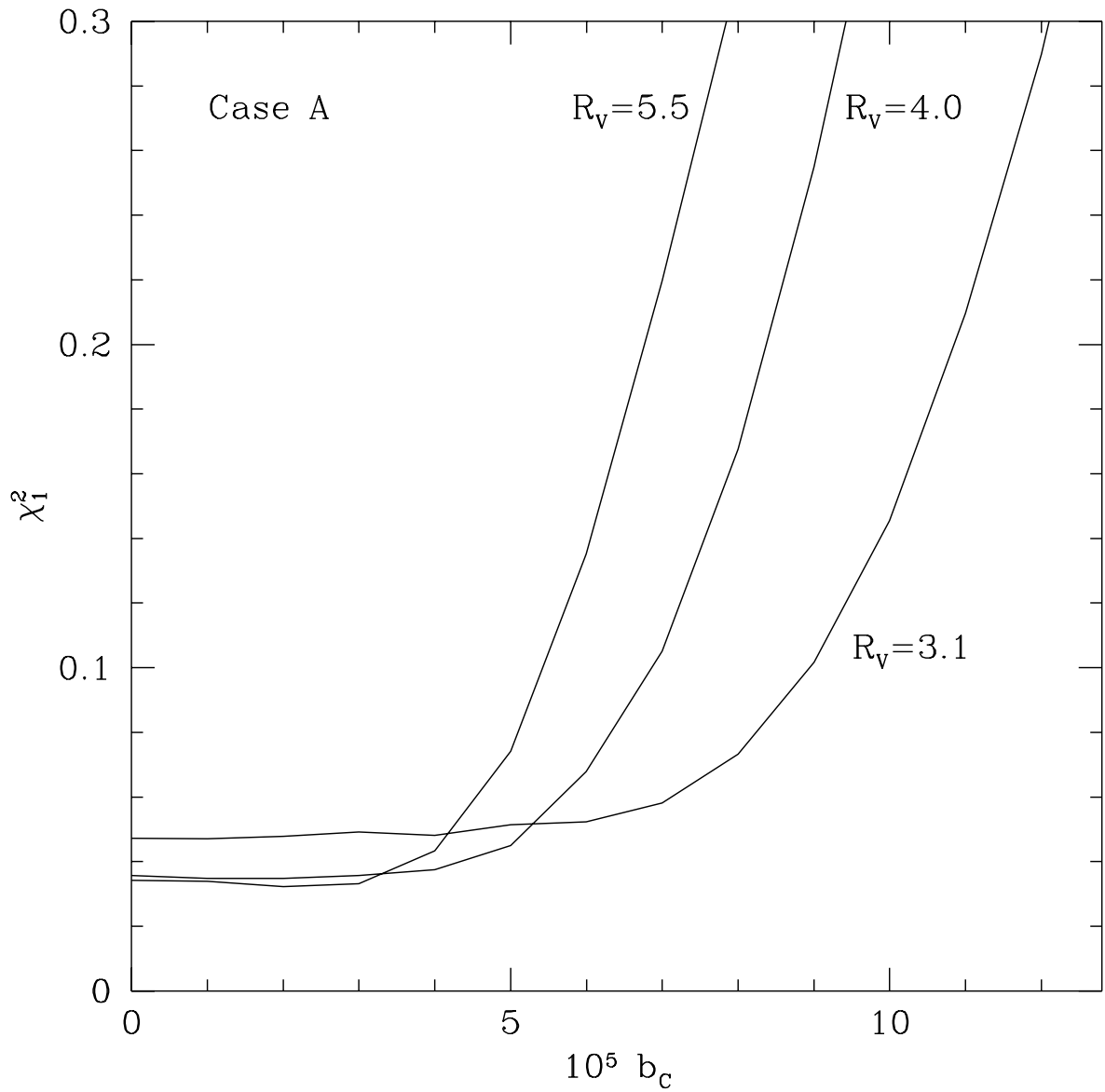}
\caption{
\label{fig:chisq}
The extinction fit error function $\chi_1^2$ (\S \ref{sec:soln}) as a function
of $b_{\rm C}$, the C abundance in the log-normal
grain population, for three values of $R_V$.  
        }
\end{figure}

In Table \ref{tab:grdistpars} we also display $\chithree^2$, 
$\chi_1^2$, and $\chi_2^2 = \sum_i (\ln A_{\rm obs} - \ln A_{\rm mod})^2$.
For a given value of $R_V$, the error functions do not vary substantially 
with $b_{\rm C}$ until a critical value of $b_{\rm C}$ is reached, at
which point the error functions increase dramatically (see Figure
\ref{fig:chisq}).
Clearly, extinction evidence alone does not constrain $b_{\rm C}$ well
except that $b_{\rm C}\ltsim 6 \times 10^{-5}$ for the $R_V=3.1$ extinction 
law, $b_{\rm C}\ltsim 4 \times 10^{-5}$ for $R_V=4$, 
and $b_{\rm C}\ltsim 3 \times 10^{-5}$ for $R_V=5.5$.
In each case, the upper limit on $b_{\rm C}$ is reached when the
very small carbonaceous particles account for 100\% of the
2175\AA\ extinction feature.

In assessing the quality of the extinction
fits, one must bear in mind that 
(1) the dielectric functions used are certainly not correct in detail,
even for bulk material, (2) the surface monolayers of grains are likely to
differ from bulk materials, (3) the true size distributions undoubtedly 
differ from the adopted functional form, and 
(4) the interstellar grains are appreciably non-spherical.  
Therefore, a precise fit is not to be expected.
One should also remember that the adopted PAH absorption cross section 
in the vacuum ultraviolet was
constructed to fit the interstellar 2175$\Angstrom$ profile, and
the silicate dielectric function in the vacuum ultraviolet was modified
to suppress structure not present in the observed interstellar extinction.

\subsubsection{Further Results}

\begin{figure}
\epsscale{1.00}
\plotone{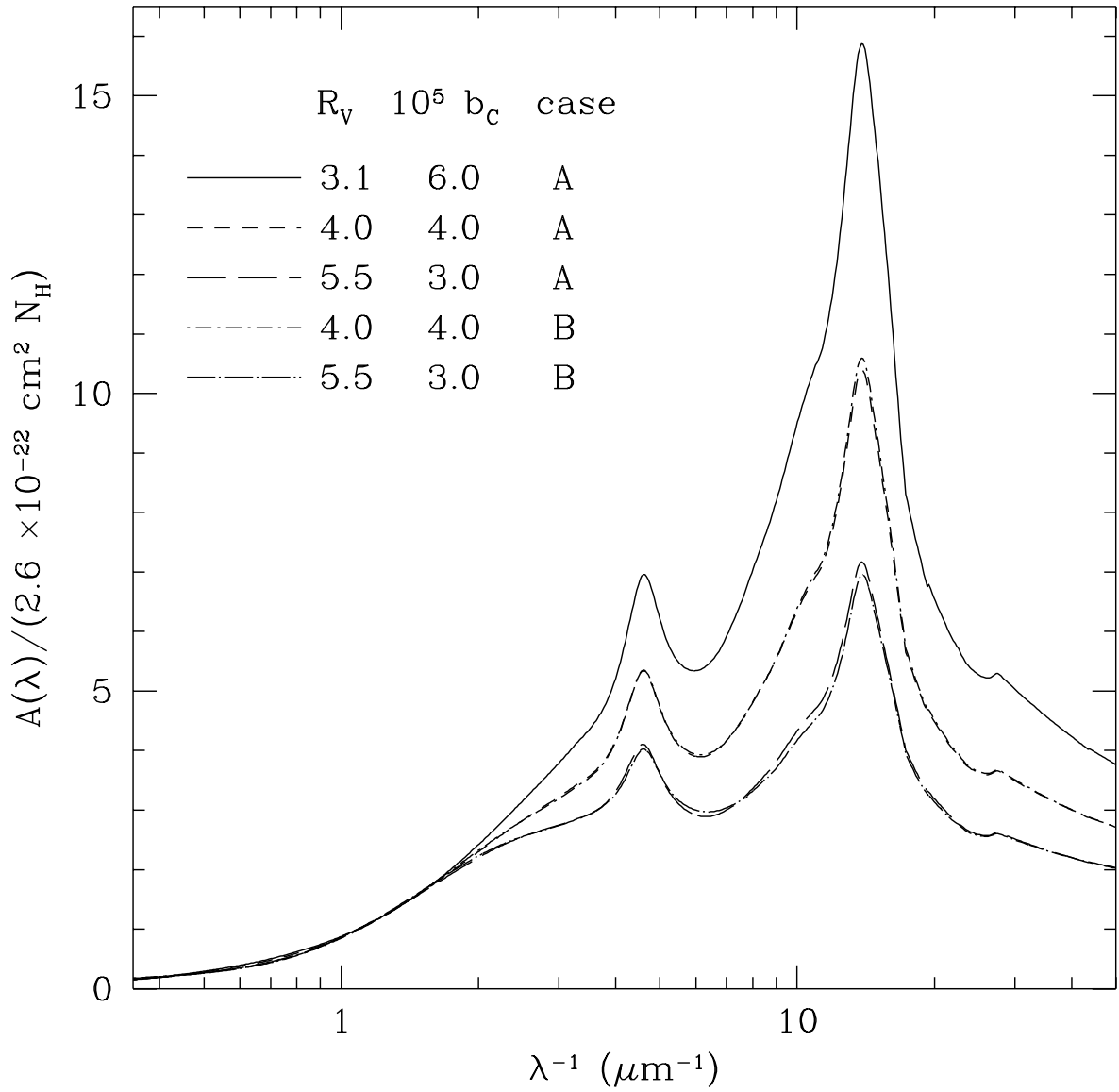}
\caption{
\label{fig:ext_shortw}
Model extinction curves extended to short wavelengths, for various size
distributions.  
        }
\end{figure}

Although neutral H gas is opaque for wavelengths
shortward of the Lyman limit,
extinction by dust at such wavelengths could have important observational 
consequences within ionized regions, including 
objects at high redshift.  Thus, in Figure 
\ref{fig:ext_shortw} we plot the model extinction resulting from several 
of our distributions over an extended wavelength range. 

\begin{figure}
\epsscale{1.00}
\plotone{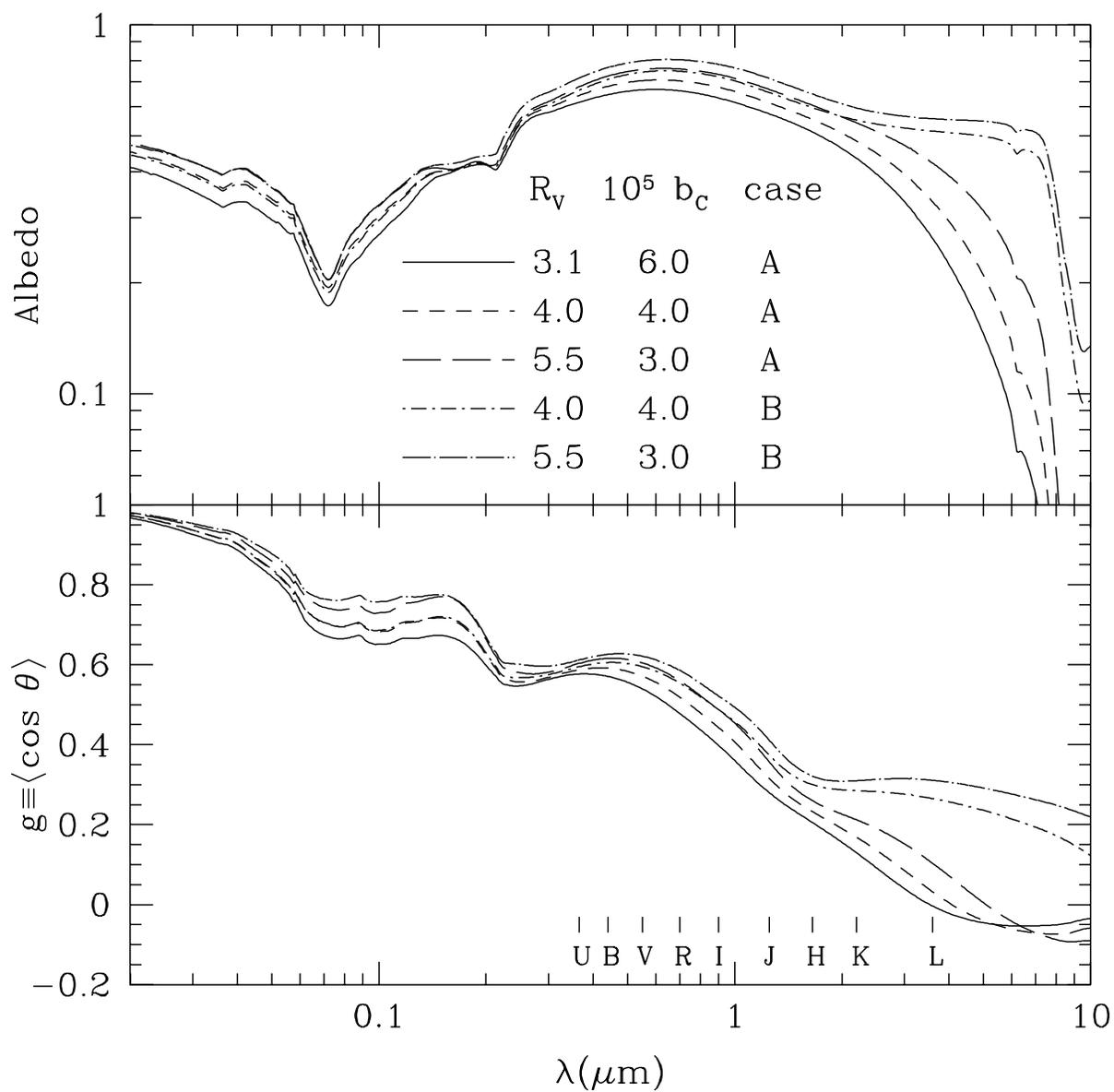}
\caption{
\label{fig:albedo}
Albedo and asymmetry parameter $g\equiv \langle \cos \theta \rangle$ for 
various size distributions.
        }
\end{figure}

In Figure \ref{fig:albedo}, we plot the albedo and asymmetry parameter
$g\equiv \langle \cos \theta \rangle$ (i.e.~the average value of $\cos 
\theta$, where $\theta$ is the angle through which 
radiation is scattered by dust) resulting from several of our model 
size distributions. 

Since Li \& Draine (2001) find that the IR emission from dust in the 
diffuse ISM is best fit when $b_{\rm C} \approx 6 \times 10^{-5}$, we
adopt this value for the $R_V=3.1$ curves in Figures \ref{fig:ext_shortw}
and \ref{fig:albedo}.  For such a large $b_{\rm C}$, the $2175 \Angstrom$
hump is almost entirely due to the very small carbonaceous
grain population.  If this
is the case for the diffuse ISM, then it seems plausible that it also 
holds in denser regions; i.e., the decrease in the strength of the 
$2175 \Angstrom$ feature with $R_V$ might result entirely from the 
depletion of very small carbonaceous grains.  Thus, we have also adopted
the large-$b_{\rm C}$ distributions for $R_V=4.0$ and 5.5 in Figures
\ref{fig:ext_shortw} and \ref{fig:albedo}.  

\subsubsection{Dust Along the Line of Sight to HD 210121}

Although the variation of the extinction curve with interstellar 
environment is fairly well characterized by the CCM parameterization, 
there are lines of sight for which the extinction deviates substantially
from CCM.  As a further test of the bare carbonaceous/silicate dust model,
it is important to seek size distributions which can reproduce the 
extinction along such sightlines.  
The extinction observed toward HD 210121 (a sightline
passing through a high-latitude diffuse molecular cloud) has
(1) an extremely small value of $R_V=2.1$, (2) a $2175\Angstrom$
feature weaker than predicted by the CCM parameterization,
and (3) a stronger-than-expected far-UV rise (see Figure 1 in
Larson et al.\ 2000).
This sightline therefore provides an opportunity to test 
the carbonaceous/silicate model and the functional forms used for our
size distributions.

\begin{figure}
\epsscale{1.00}
\plotone{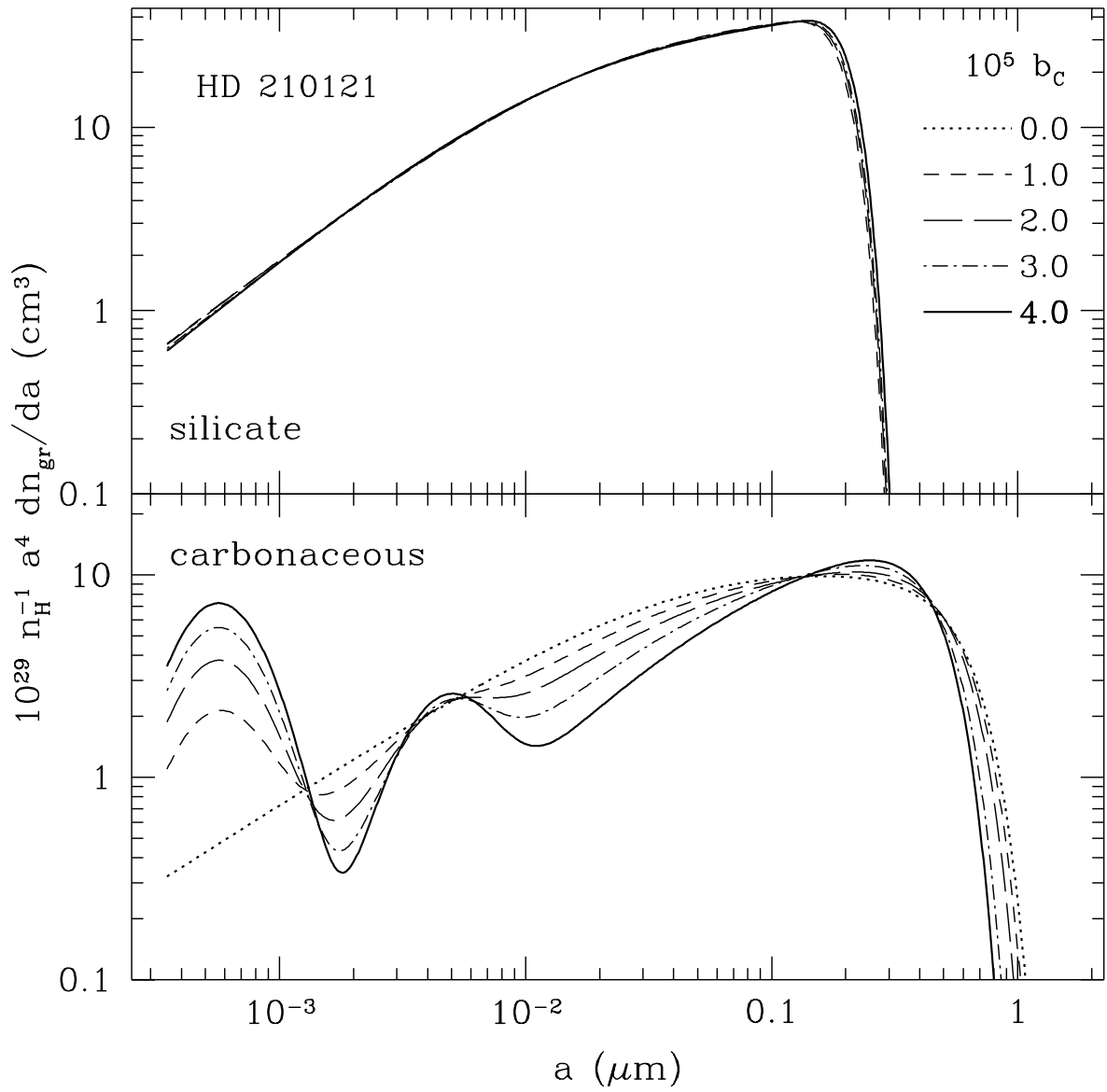}
\caption{
\label{fig:grdist_HD210121}
Grain size distributions for HD 210121.  
	}
\end{figure}

\begin{figure}
\epsscale{1.00}
\plotone{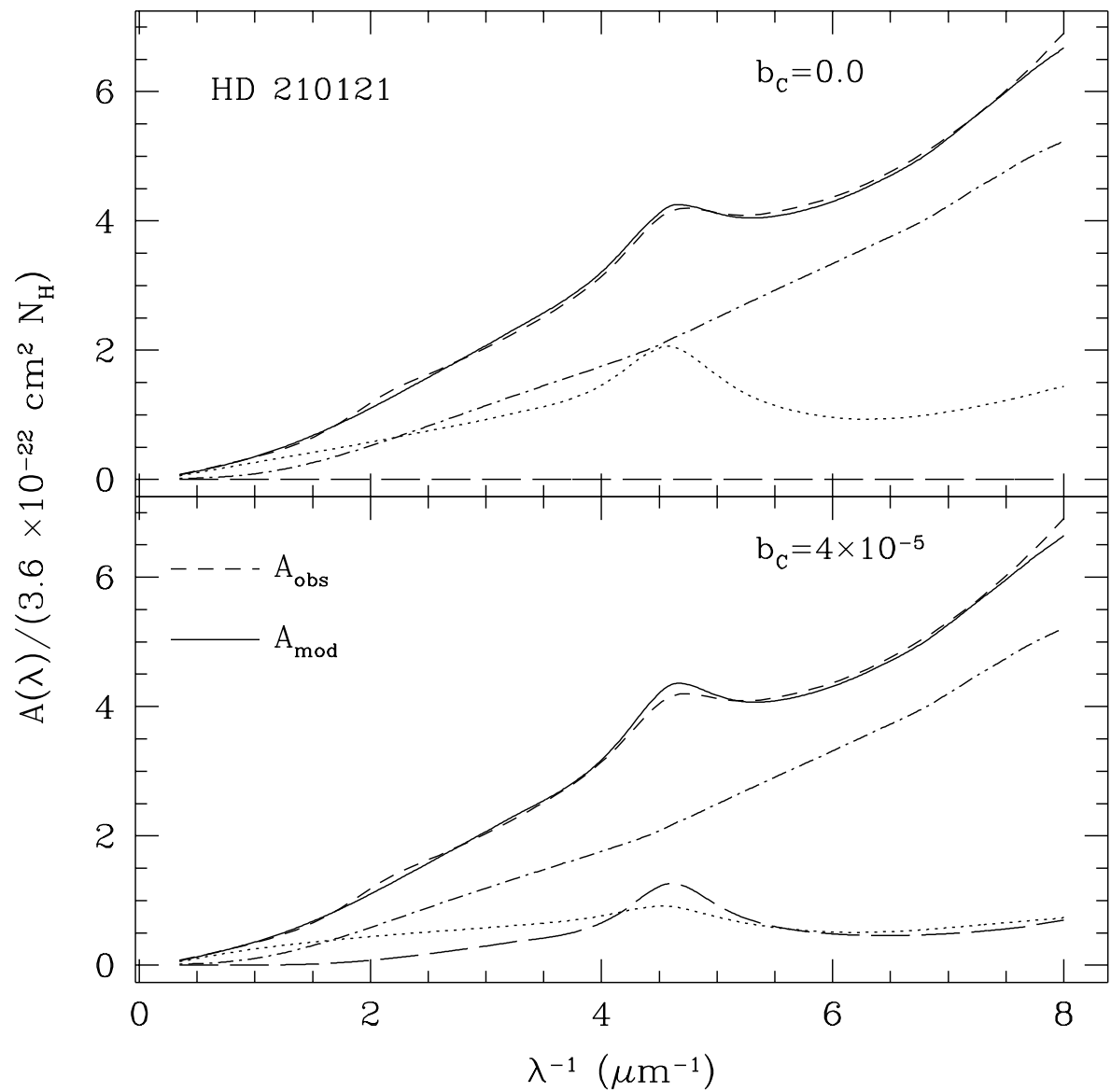}
\caption{
\label{fig:ext_HD210121}
Same as Figure \ref{fig:ext_3.1}, but for the extinction along the line of
sight to HD 210121 
and $b_{\rm C} = 0.0$ and $4.0 \times 10^{-5}$.  Note the difference in 
vertical scale from Figure \ref{fig:ext_3.1}.
        }
\end{figure}

Larson et al.\ (2000) used the maximum entropy method to construct
size distributions for the grains toward HD 210121.
We seek to reproduce the extinction toward HD 210121 (Larson et al.~2000;
Larson, Whittet, \& Hough 1996; Welty \& Fowler 1992) with size distributions
of our simple functional form.  We adopt the normalization given by 
Larson et al.~(2000):  $A_V/N_{\rm H} = 3.6 \times 10^{-22} \cm^2$.  In 
fitting the extinction, we adopt 100 points equally spaced in $\lambda^{-1}$
rather than in $\ln \lambda$.  We have found that this yields a better fit
to the $2175 \Angstrom$ hump and far-UV rise without compromising the fit
quality in the infrared.  
Distribution parameter values are given in Table \ref{tab:HD210121} and
the distributions and extinction fits are plotted in Figures 
\ref{fig:grdist_HD210121} and \ref{fig:ext_HD210121}, respectively.

We are able to obtain acceptable fits to the extinction toward HD 210121
with values of $b_c$ ranging up to $4\times10^{-5}$, and reasonable size
distributions for the carbonaceous and silicate grain populations.
Our grain model successfully accomodates this line of sight with its 
extremely small value of $R_V$ and deviation from the CCM parameterization.

\subsection{Dust in the Magellanic Clouds}

The metallicities in the Magellanic Clouds are substantially lower than in
the Milky Way, and measured extinction curves toward stars
in the LMC and SMC differ from typical extinction curves in the
Milky Way.
The LMC and SMC therefore offer opportunities to test the applicability of
our grain model to low-metallicity extragalactic 
environments.\footnote{See Pei (1992) for an early extension of the MRN 
model to the Magellanic Clouds.}

Clayton et al.~(2000) used the maximum 
entropy method to find graphite/silicate size distributions that accurately
reproduce the extinction along various Magellanic Cloud sightlines.  Here,
we seek distributions of our simple functional form that reproduce the 
average extinction 
in the LMC (Misselt, Clayton, \& Gordon 1999), 
the extinction in the LMC 2 area (Misselt et al.~1999), 
and the extinction in the 
SMC bar, along the line of sight to the star AzV398 (Gordon \& Clayton 1998).  
For $\lambda^{-1} \ltsim 3 \micron^{-1}$, the extinction is determined at 
only a small number of wavelengths.  Thus, for the Magellanic Clouds, we  
evaluate the extinction at 100 wavelengths spaced equally in $\lambda^{-1}$, 
rather than in $\ln \lambda$.  

The extinction normalization and elemental abundances are even more uncertain
for the Magellanic clouds than for the Milky Way.  For the LMC,
Koorneef (1982) found $N({\rm H \, I})/E(B-V) = 2.0 \times 10^{22} \cm^{-2}$
and Fitzpatrick (1985) found 
$N({\rm H \, I})/E(B-V) = 2.4 \times 10^{22} \cm^{-2}$.  Averaging these
results and taking 
$R_V=2.6$ (the average for the 10 measured $R_V$ values in
Misselt et al.'s sample), 
we adopt $A(V)/N_{\rm H} = 1.2 \times 10^{-22}
\cm^2$.  For the SMC, Martin, Maurice, \& Lequeux (1989) found
$N_{\rm H}/E(B-V) = 4.6 \times 10^{22} \cm^{-2}$; with $R_V = 2.87$ 
(Gordon \& Clayton 1998)
this yields $A(V)/N_{\rm H} = 6.2 \times 10^{-23} \cm^2$.  We take 
the abundance/depletion-limited values of 
$V_{\rm tot,g}$ and $V_{\rm tot,s}$ to be reduced from their values in the 
Milky Way by a factor 1.6 for the LMC and 4.0 for the SMC (Gordon \& 
Clayton 1998).   

\begin{figure}
\epsscale{1.00}
\plotone{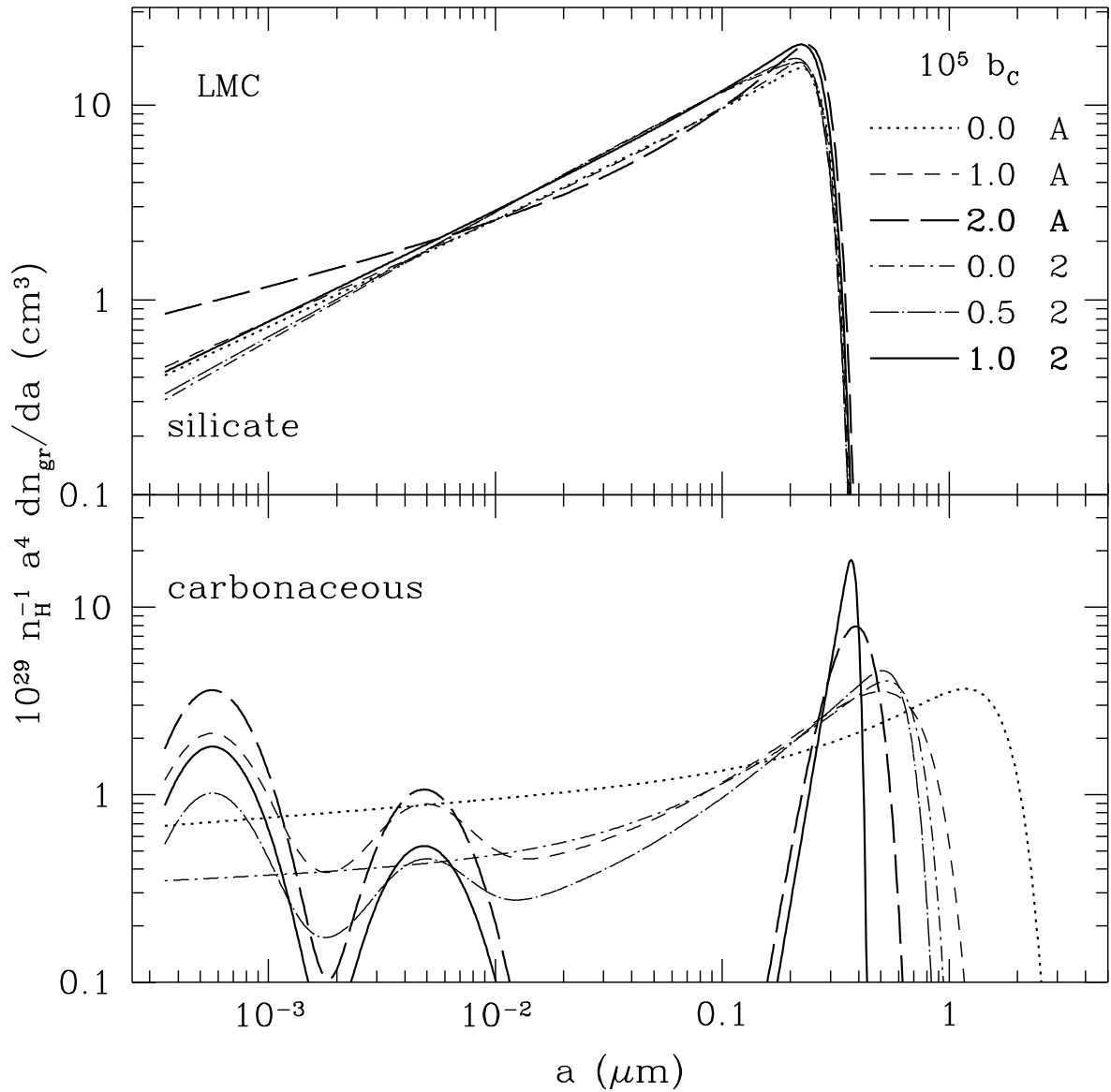}
\caption{
\label{fig:grdist_LMC}
Grain size distributions for the LMC.  The values of $b_{\rm C}$ are 
indicated; ``A'' denotes distributions constructed to fit the average 
extinction in the LMC and ``2''denotes distributions for the LMC 2 area.
	}
\end{figure}

\begin{figure}
\epsscale{1.00}
\plotone{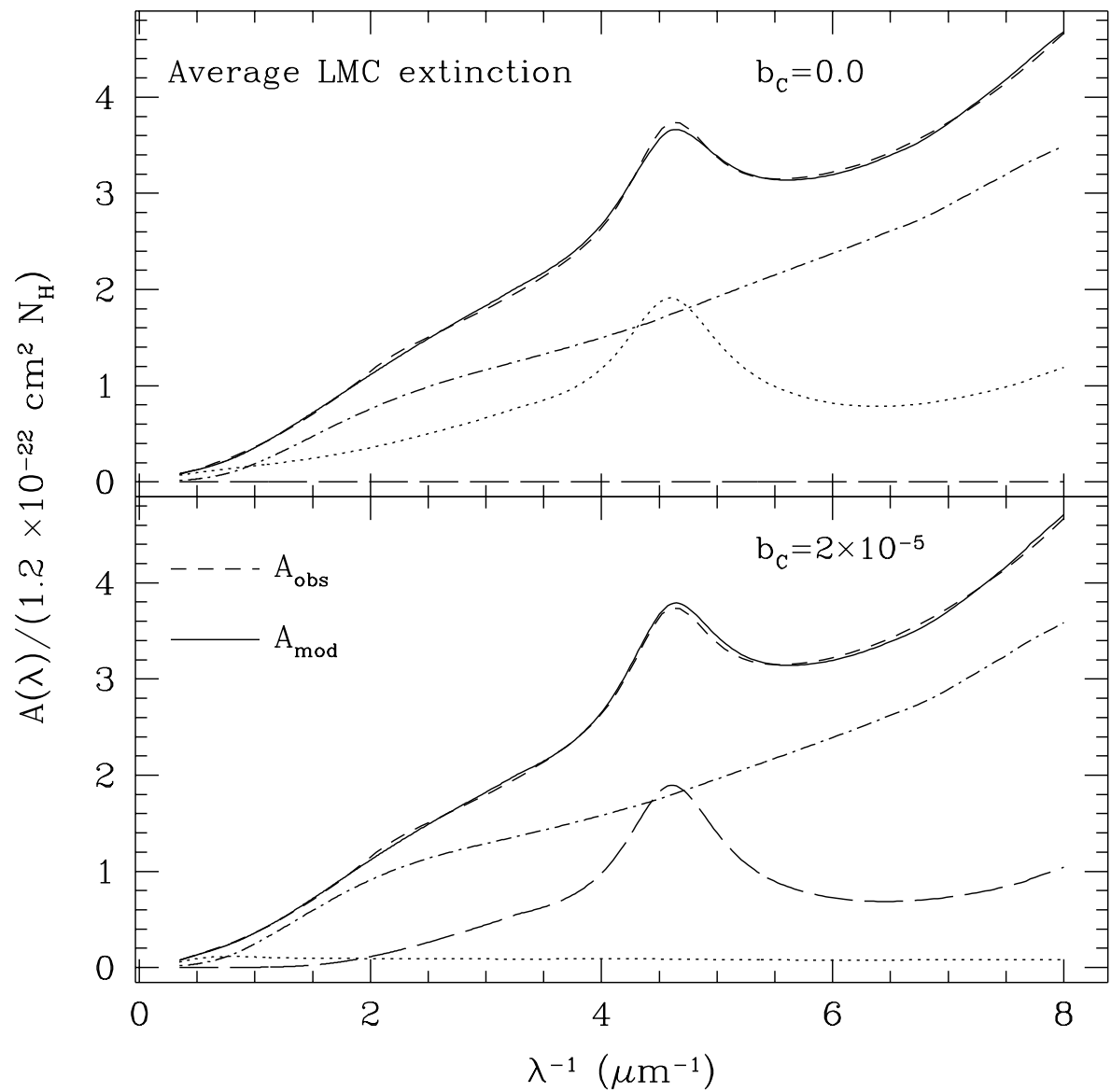}
\caption{
\label{fig:ext_LMCavg}
Same as Figure \ref{fig:ext_3.1}, but for the average extinction for the LMC 
and $b_{\rm C} = 0.0$ and $2.0 \times 10^{-5}$.  Note the difference in 
vertical scale from Figure \ref{fig:ext_3.1}.
        }
\end{figure}

\begin{figure}
\epsscale{1.00}
\plotone{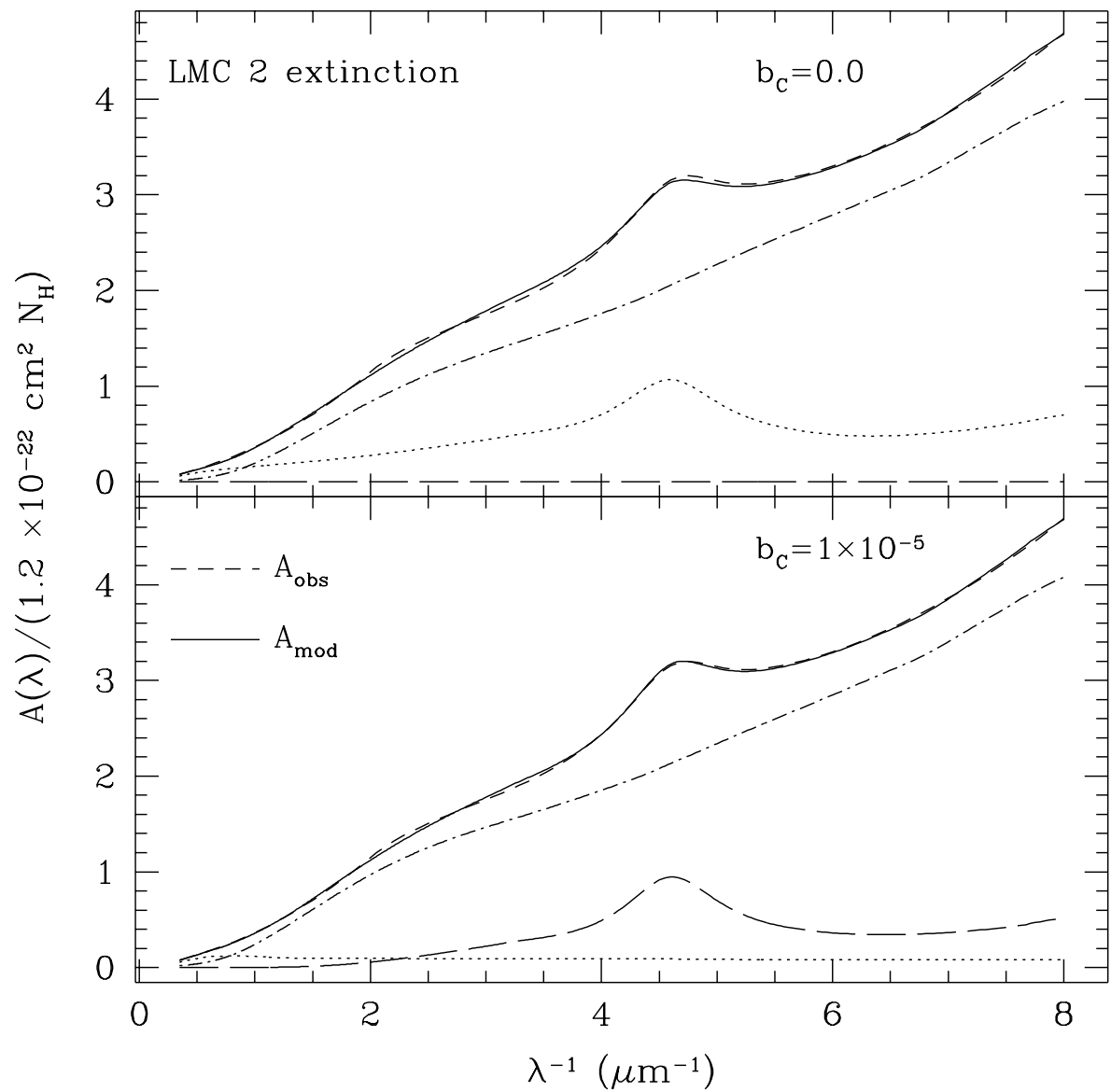}
\caption{
\label{fig:ext_LMC2}
Same as Figure \ref{fig:ext_3.1}, but for the LMC 2 area 
and $b_{\rm C} = 0.0$ and $1.0 \times 10^{-5}$.  Note the difference in
vertical scale from Figure \ref{fig:ext_3.1}.
        }
\end{figure}

\begin{figure}
\epsscale{1.00}
\plotone{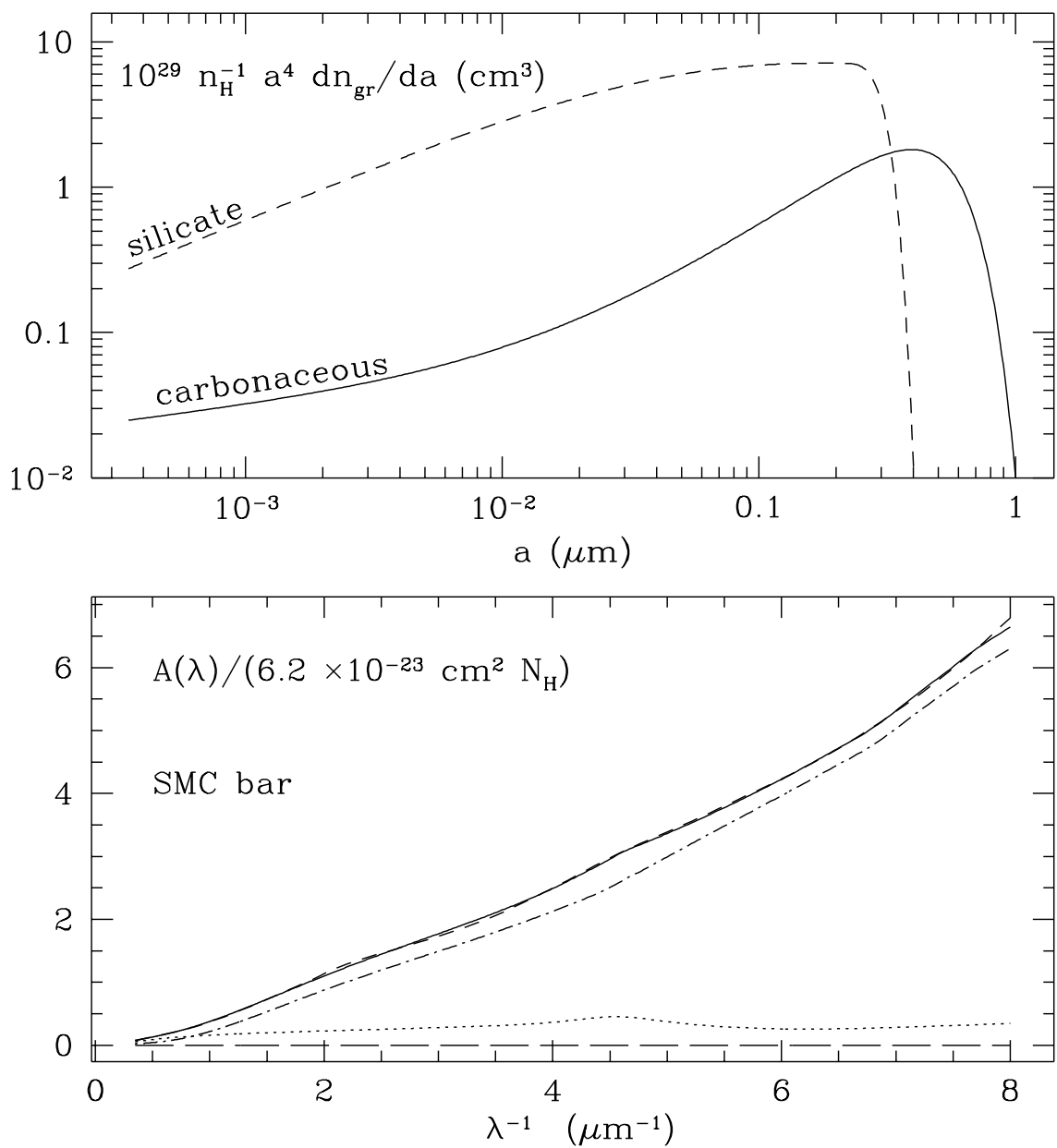}
\caption{
\label{fig:SMCbar}
Upper panel:  Size distribution for the SMC bar, with $b_{\rm C}=0.0$.
Lower panel:  The corresponding extinction fit; curve types are the same as
in Figure \ref{fig:ext_3.1}.
        }
\end{figure}

\begin{figure}
\epsscale{1.00}
\plotone{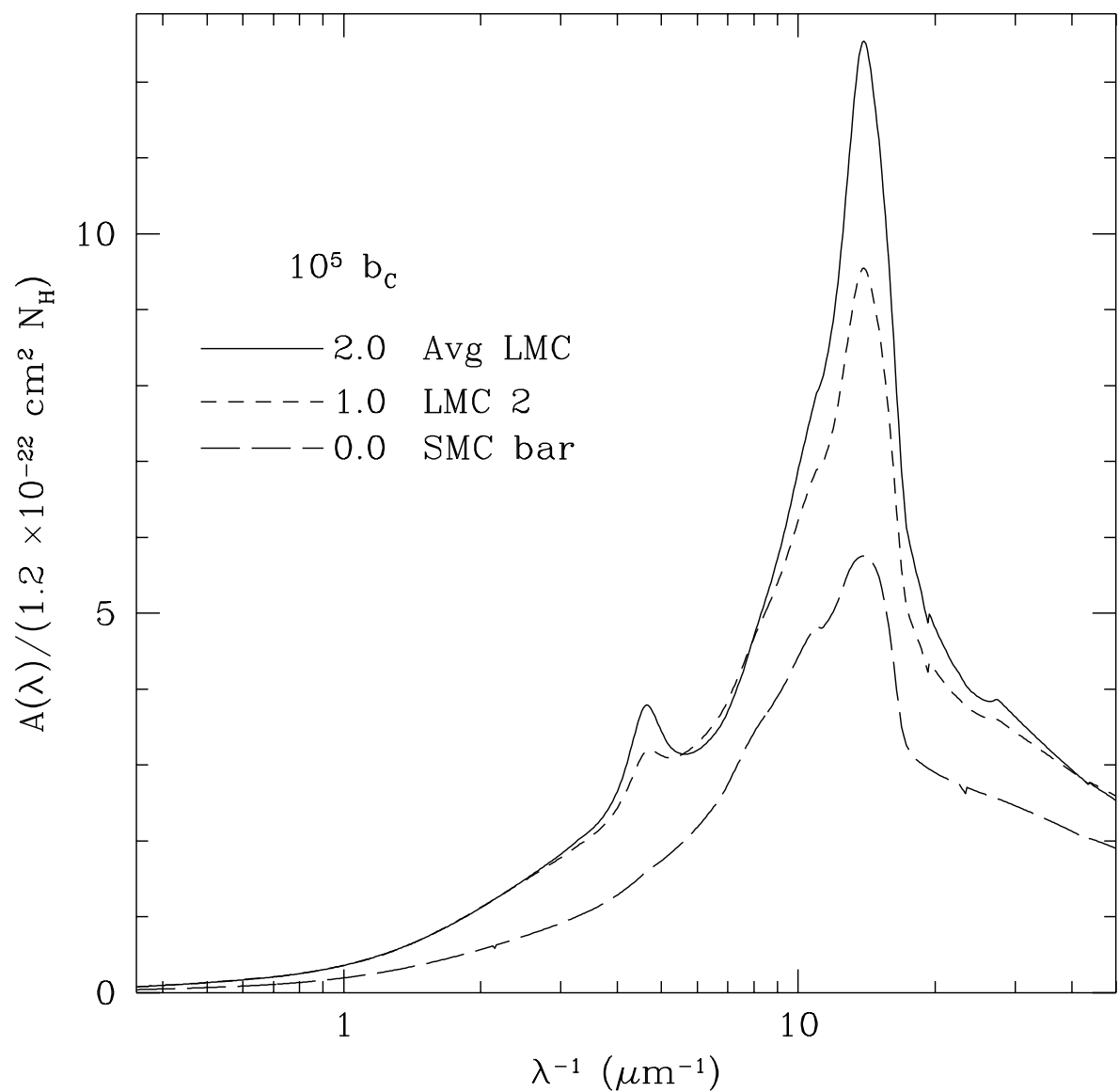}
\caption{
\label{fig:ext_shortw_MC}
Model extinction curves extended to short wavelengths, for Magellanic 
Cloud environments.
        }
\end{figure}

\begin{figure}
\epsscale{1.00}
\plotone{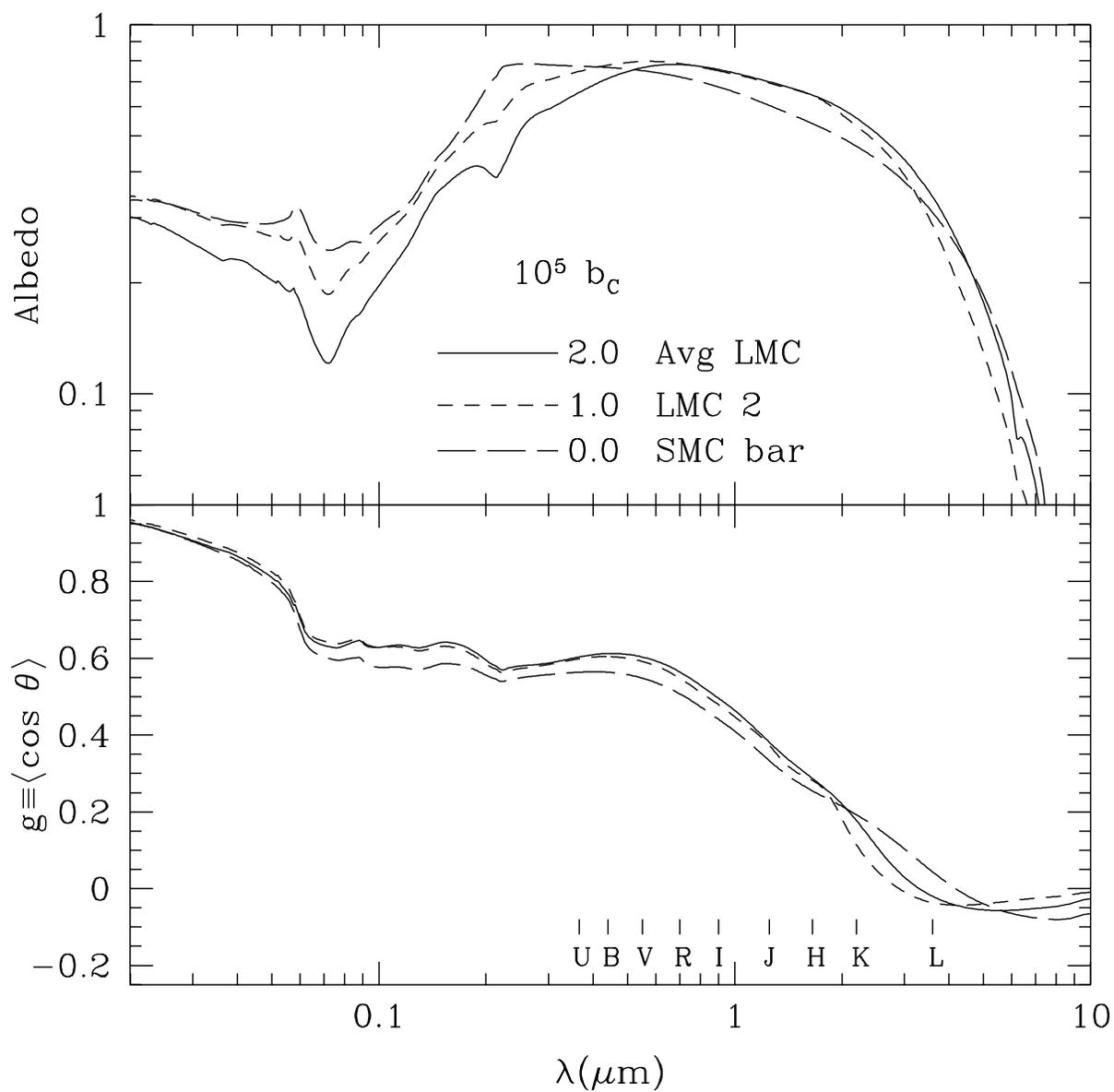}
\caption{
\label{fig:albedo_MC}
Albedo and asymmetry parameter $g\equiv \langle \cos \theta \rangle$ for 
Magellanic Cloud environments.
        }
\end{figure}

Distribution parameters for which the extinction is best fit are given in 
Table \ref{tab:grdistpars_MC}.  We also tabulate the total grain volumes, 
normalized to the limiting values estimated in the previous paragraph; note
that all of the LMC distributions 
use less than the estimated available amount
of C and Si.  Size distributions, extinction fits, and related quantities 
are plotted in Figures \ref{fig:grdist_LMC} through \ref{fig:albedo_MC}.

Note the absence of the $2175 \Angstrom$ feature in the SMC bar extinction
curve (Figure \ref{fig:SMCbar}), which implies the absence of very small 
carbonaceous grains.  Recently, Reach et al.~(2000) have detected PAH 
emission features in a quiescent molecular cloud in the SMC.  Reach et 
al.~point out that SMC extinction curve measurements are biased towards hot,
luminous stars, so that very small grains may have been destroyed along 
these sightlines.

\section{Discussion}

\subsection{Abundances and Grain Models}

Note from Table \ref{tab:grdistpars} that, in the Milky Way, 
the silicate volumes generally exceed the 
abundance/depletion-limited value, by $\approx 10\%$ when $R_V=5.5$ to 
$\approx 30 \%$ when $R_V=3.1$, and the carbonaceous grain volume exceeds
its abundance/depletion-limited value by $\approx 10\%$ when $R_V=3.1$.  
We would expect non-spherical grains to produce more extinction per unit 
grain volume than spheres, so that our violation of abundance 
constraints might be an artifact due to the use of only spherical grains in
our modelling.  However, we have used the discrete dipole approximation 
(Draine \& Flatau 1994; Draine 2000)
to calculate extinction efficiencies for silicate grains of various shapes
with $a \ge 0.01 \micron$ and have found that the
integrated extinction per grain volume, $\int (C_{\rm ext}/V)d\lambda$
integrated over $\lambda^{-1} \in [0.35, 8.0] \micron^{-1}$, varies
only slightly with shape.   

Kim et al.~(1994) sought to maximize the efficient use of grain volume by 
allowing more complicated size distributions.  Although such an approach
could lower the total amount of grain volume that we need to reproduce the
observed extinction, we find such fine-tuning unappealing.  
It seems to us unlikely that nature has produced size distributions
fine-tuned to maximize the extinction per volume over just the
wavelengths where we are able to measure the extinction.
We think it more likely that either
the true elemental abundances in the ISM really are somewhat
higher than in the Sun,
or that the bare graphite/silicate model is inadequate in some more 
fundamental way.

Other well-developed models include composite, fluffy grains 
(Mathis 1996, 1998)
and grains consisting of silicate cores covered by organic refractory 
mantles (Li \& Greenberg 1997).  The recent discovery that
the $3.4 \micron$ aliphatic C-H stretch absorption feature 
toward Sgr A IRS7 
is unpolarized (whereas the $9.7 \micron$ silicate absorption feature 
toward Sgr A IRS3 
is polarized) may rule out the core-mantle model (Adamson et al.~1999), 
although model 
calculations of the relative polarization in these features have not yet 
been carried out 
for the core-mantle model, and the silicate feature polarization has
yet to be measured for IRS 3 itself.

Mathis (1996) found that a
mixture of composite grains (consisting of small silicate and amorphous 
carbon grains and $\approx 45\%$ vacuum), small graphite grains, and some
small silicate grains could reproduce the observed extinction while 
incorporating C, Si, Fe, and Mg with substantially sub-solar abundances.
However, there are some difficulties with this model.  First, Mathis 
adopts dielectric functions for the composite grains using effective medium
theory, calculates extinction cross sections for spheres, and then 
multiplies the cross sections by a factor 1.09, to account for enhancements
in extinction due to non-spherical shapes.  The final step must be viewed
with suspicion, since it fails for compact silicate grains.  

Also, 
Mathis used the optical properties of ``Be'' amorphous carbon from Rouleau
\& Martin (1991).  Schnaiter et al.~(1998) have pointed out that the 
derived optical properties, while possibly correct, are unproven, since the
adopted description of the sample geometry as a continuous distribution of
ellipsoids is so simplistic that substantial errors can result.  
Furthermore, ``Be'' amorphous carbon is much more absorbing at long 
wavelengths than various forms of hydrogenated amorphous carbon, and this 
absorption provides most of 
the extinction for $\lambda^{-1} \ltsim 3 \micron^{-1}$
in the Mathis (1996) composite model.\footnote{Dwek (1997) has argued that
	the fluffy grain model employing ``Be'' amorphous carbon produces
	too much IR emission compared with the COBE data (Dwek et al.~1997).}  
Furton, Laiho, \& Witt (1999) have performed 
laboratory studies of hydrogenated amorphous carbon, and find that such
grains can reproduce the observed $3.4 \micron$ absorption feature if the 
degree of hydrogenation is rather large ($\approx 0.5 \,$H/C).  
There is very little visible/IR continuum
absorption in this case.  Thus, the composite
model does not simultaneously provide enough long-wavelength extinction
and $3.4 \micron$ absorption.  Of course, the bare graphite/silicate model
does not account for the $3.4 \micron$ absorption 
either.\footnote{To accomodate the 3.4$\micron$ feature, the graphite/silicate
	model must be extended to include 
	aliphatic hydrocarbons, possibly within hydrogenated carbon coatings
	on the large graphitic grains.
	}

Although the bare graphite/silicate model apparently requires higher 
abundances of C, Si, Fe, and Mg than are generally thought to be 
available in the ISM, it would be premature to abandon it.  The true 
interstellar abundances are not yet known, and the 
alternatives have difficulties too.  
Further progress in dust modelling will require the determination of
dielectric functions for amorphous carbons with a range of degrees of 
hydrogenation, over the full range $\lambda^{-1} \in [0.35, 8.0] \micron^{-1}$,
as well as detailed modelling of how the extinction per unit volume varies
depending on grain geometry.

\subsection{Observed Size Distribution of Interstellar Grains Streaming 
Through the Solar System}

\begin{figure}
\epsscale{1.00}
\plotone{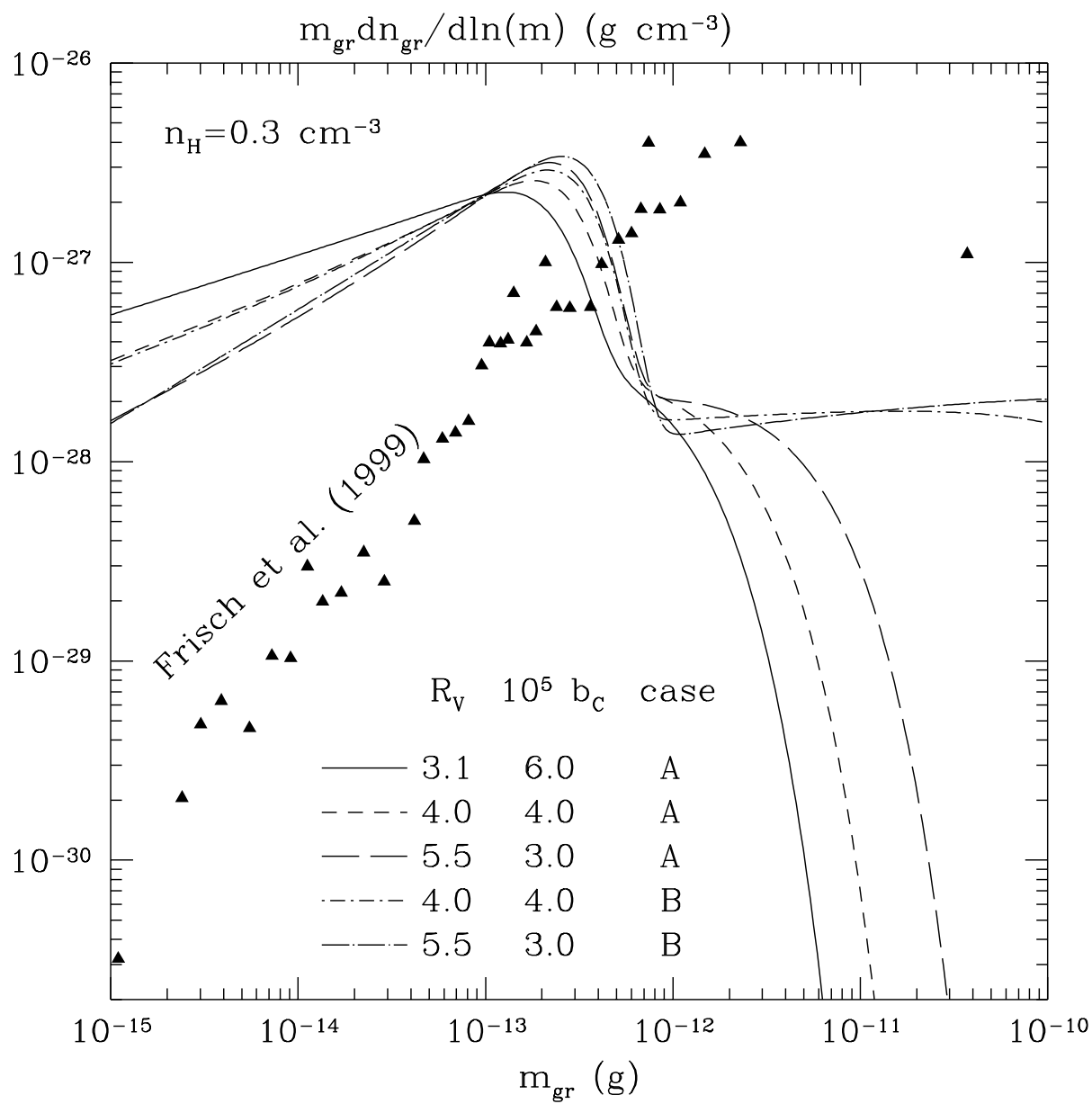}
\caption{
\label{fig:frisch}
Mass distribution for grains in the local ISM determined by Frisch et 
al.~(1999) (triangles).  Mass distributions for the size distributions
of \S \ref{sec:results} are also shown; the sharp drop at $m \sim
3 \times 10^{-13} \, {\rm g}$ corresponds to the rapid drop in silicate
grain abundance at $a \sim 0.3 \micron$.
        }
\end{figure}

Recently, Frisch et al.~(1999) have presented a grain mass distribution
for the local interstellar medium (LISM), derived from the measured 
rate of impact of interstellar grains with
detectors on the {\it Ulysses} and {\it Galileo} spacecraft;  we reproduce
their data points in Figure \ref{fig:frisch}.  We also show mass 
distributions as derived here from fitting extinction, for $(R_V, 10^5
b_{\rm C})=$ (3.1, 3.0), (4.0, 2.0), and (5.5, 1.0).
We adopt $\nH = 0.3 \cm^{-3}$, as 
recommended by Frisch et al.  Note that none of our distributions resemble
the Frisch et al.~result.  The steep drop in the Frisch et al.~distribution
at small masses probably reflects the exclusion of small grains from the
solar system 
-- smaller grains are more tightly coupled to the magnetic field
and are less likely to penetrate the heliosphere to within $\sim 5 \, 
\rm{AU}$ of the Sun (Linde \& Gombosi 2000).
However, the large amount of mass in large grains
in the Frisch et al.~distribution is hard to fathom.  The error bars on the
Frisch et al.~data (not shown in Figure \ref{fig:frisch}) are large; 
further observations of interstellar dust entering the solar system
would be of great value.

If the Frisch et al.\ result is confirmed, then there are two possibilities.
If the region through which the solar system is now passing contains
a truly representative dust-gas mixture, then a dramatically different
grain model would be required.  It is difficult to envision a grain
model which could simultaneously account for the interstellar extinction law,
be consistent with interstellar elemental abundances, and reproduce the
Frisch et al.\ size distribution.
Alternatively, it could be the case that size-sorting and gas-grain
separation occur on small scales in the ISM, and that the region through
which the solar system is now moving happens to have an unusual
concentration of large grains.

\subsection{Conclusions}

The simplest interstellar dust model consists of a population of 
carbonaceous grains
and a separate population of silicate grains.  In the original development
of this model by MRN, the grain size distribution was chosen so as to 
reproduce the observed extinction for lines of sight with $R_V \approx 3.1$.
The observation of relatively short-wavelength infrared
emission from dust implies that there are 
substantial numbers of very small (mainly carbonaceous) grains,
smaller than the lower cutoff size of the 
MRN distribution.  Furthermore, the extinction curve has been found to vary
substantially depending on the interstellar environment through which the
starlight passes; thus, there is no single grain size distribution which
applies in all environments.  By finding carbonaceous/silicate grain size
distributions which contain sufficient very small grains to account
for the observed infrared emission (Li \& Draine 2001),
and which reproduce the observed extinction for 
a wide range of environments, we have demonstrated that the simplest dust 
model remains viable.  

Although difficulties remain, they are no more 
severe than the difficulties with other, more complicated, models.
These difficulties include the requirement of somewhat super-solar 
abundances of the dust constituent elements, the lack of a $3.4 \micron$
absorption feature in a model in which all of the C is in graphite or 
PAHs, and the gross disparity between the derived grain size distributions
and that inferred by Frisch et al.~(1999) for dust in the local ISM.  
Additionally, there is evidence from depletion patterns that 
metallic Fe or Fe oxides are an important dust component
(Sofia et al.~1994; Howk et al.~1999).
The observed 90 GHz emission from interstellar dust appears to
rule out a substantial metallic Fe component (Draine \& Lazarian 1999),
but oxides such as FeO or magnetite Fe$_3$O$_4$ are not excluded.
Dielectric functions for candidate Fe oxides are needed to investigate
such grain models.

Finally,
the variation in the grain size distribution with environment seems
to indicate that small grains coagulate onto large grains in relatively
dense environments, as expected (Draine 1985; Draine 1990).
Presumably, mass is returned from large to small grains
via shattering during grain-grain collisions in 
shock waves.  (Mass is also returned to the gas via sputtering processes.)
Weingartner \& Draine (1999) found that the observed elemental 
depletions in the interstellar medium could be due to accretion onto grains
if the timescales for matter to cycle between interstellar phases are
$\sim 10^7 \yr$.  
It remains a 
mystery how two separate grain 
populations -- carbonaceous grains and silicate grains -- could remain 
distinct after evolving through many cycles of coagulation, shattering, 
accretion, and erosion; perhaps they do not.

While real grains are undoubtedly more complex,
the graphite/silicate model for dust in diffuse clouds is clearly-defined,
and consistent with observations of
interstellar extinction in the Milky Way, LMC, and SMC
(as demonstrated in the present work)
and infrared emission (Li \& Draine 2001).
While the model 
does not explicitly account for the $3.4\micron$ feature or the relatively
weak diffuse interstellar bands (Herbig 1995), these could
conceivably be accomodated 
by modest modifications of or extensions to
the basic graphite/silicate model.
The ``extended red emission'' from interstellar dust (Witt \& Boroson 1990)
could also perhaps be due to a minor
modification of the basic graphite/silicate model (e.g., a hydrogenated
amorphous carbon coating; Witt \& Furton 1995).

Until a more compelling grain model is available,
we recommend the use of 
the simplest one, specified by the size distributions found here and 
optical properties given by Draine \& Lee (1984), Laor \& Draine (1993),
and Li \& Draine (2001).  In particular, we favor the distributions
with relatively large $b_{\rm C}$ (Li \& Draine 2001), for which the very
small carbonaceous grain population entirely accounts for the 
$2175 \Angstrom$ hump in the extinction curve.  

\acknowledgements
This research was supported in part by NSF grant
AST-9619429 and by NSF Graduate and International Research Fellowships to JCW.
We are grateful to Eli Dwek, 
Aigen Li, and John Mathis for helpful discussions and to 
R. H. Lupton for the availability of the SM plotting 
package.

Note added in proof---Li \& Draine (2000) have recently found that the 
nondetection of the $10 \micron$ silicate feature in emission from diffuse
clouds does not strongly constrain the ultrasmall silicate grain population,
since the $10 \micron$ feature may be hidden by the dominant PAH features.
Li \& Draine estimate that as much as $\sim 20\%$ of the interstellar Si 
could be in grains with $a\ltsim 15 \Angstrom$.


\clearpage


\renewcommand{\tabcolsep}{0.25cm}
\setlength{\textheight}{9.5in}
\newpage
\begin{deluxetable}{ccccccccccccccccc}
\tabletypesize{\scriptsize}
\rotate
\tablewidth{0pt}
\tablecaption{Grain Size Distribution Parameter 
Values\tablenotemark{a} 
\label{tab:grdistpars}}
\tablehead{
\colhead{$R_V$\tablenotemark{b}}&
\colhead{$10^5 b_{\rm C}$\tablenotemark{c}}&
\colhead{case}&
\colhead{$\alpha_{\rm g}$}&
\colhead{$\beta_{\rm g}$}&
\colhead{$a_{\rm t,g}$}&
\colhead{$a_{\rm c,g}$}&
\colhead{$C_{\rm g}$}&
\colhead{$\alpha_{\rm s}$}&
\colhead{$\beta_{\rm s}$}&
\colhead{$a_{\rm t,s}$}&
\colhead{$C_{\rm s}$}&
\colhead{$\tilde{V}_{\rm g}$\tablenotemark{d}}&
\colhead{$\tilde{V}_{\rm s}$\tablenotemark{d}}&
\colhead{$\chi_1^2$\tablenotemark{e}}&
\colhead{$\chi_2^2$\tablenotemark{f}}&
\colhead{$\chithree^2$\tablenotemark{g}}
\\
\colhead{}&
\colhead{}&
\colhead{}&
\colhead{}&
\colhead{}&
\colhead{($\micron$)}&
\colhead{($\micron$)}&
\colhead{}&
\colhead{}&
\colhead{}&
\colhead{($\micron$)}&
\colhead{}&
\colhead{}&
\colhead{}&
\colhead{}&
\colhead{}&
\colhead{}

}
\startdata
3.1  &0.0  &A &-2.25  &-0.0648  &0.00745   &0.606   
&$9.94 \times 10^{-11}$ &
-1.48  &-9.34  &0.172      
&$1.02 \times 10^{-12}$
&1.146  &1.244  &0.047  &0.111  &0.118 \\
3.1  &1.0  &A &-2.17  &-0.0382  &0.00373  &0.586   
&$3.79 \times 10^{-10}$ &
-1.46  &-10.3  &0.174      
&$1.09 \times 10^{-12}$
&1.137  &1.251  &0.047  &0.116  &0.118 \\
3.1  &2.0  &A &-2.04  &-0.111  &0.00828  &0.543   
&$5.57 \times 10^{-11}$ &
-1.43  &-11.7  &0.173      
&$1.27 \times 10^{-12}$
&1.130  &1.254  &0.048  &0.124  &0.118 \\ 
3.1  &3.0  &A &-1.91  &-0.125   &0.00837   &0.499   
&$4.15 \times 10^{-11}$ &
-1.41  &-11.5  &0.171     
&$1.33 \times 10^{-12}$
&1.119  &1.260  &0.049  &0.139  &0.119 \\
3.1  &4.0  &A &-1.84  &-0.132   &0.00898   &0.489   
&$2.90 \times 10^{-11}$ &
-2.10  &-0.114  &0.169      
&$1.26 \times 10^{-13}$
&1.113  &1.290  &0.048  &0.135  &0.126 \\
3.1  &5.0  &A &-1.72  &-0.322   &0.0254   &0.438   
&$3.20 \times 10^{-12}$ &
-2.10  & -0.0407  &0.166      
&$1.27 \times 10^{-13}$
&1.098  &1.304  &0.051  &0.154  &0.131 \\
3.1  &6.0  &A &-1.54  &-0.165   &0.0107   &0.428   
&$9.99 \times 10^{-12}$ &
-2.21  & 0.300  &0.164      
&$1.00 \times 10^{-13}$
&1.092  &1.322  &0.052  &0.161  &0.136 \\ 
4.0  &0.0  &A &-2.26  &-0.199  &0.0241   &0.861   
&$5.47 \times 10^{-12}$ &
-2.03   &0.668  &0.189      
&$5.20 \times 10^{-14}$
&1.000  &1.100  &0.036  &0.100  &0.048 \\
4.0  &1.0  &A &-2.16  &-0.0862  &0.00867   &0.803   
&$4.58 \times 10^{-11}$ &
-2.05  & 0.832  &0.188      
&$4.81 \times 10^{-14}$
&0.992  &1.103  &0.035  &0.104  &0.048 \\
4.0  &2.0  &A &-2.01  &-0.0973   &0.00811   &0.696   
&$3.96 \times 10^{-11}$ &
-2.06  & 0.995  &0.185      
&$4.70 \times 10^{-14}$
&0.974  &1.112  &0.035  &0.113  &0.050 \\
4.0  &3.0  &A &-1.83  &-0.175   &0.0117  &0.604   
&$1.42 \times 10^{-11}$ &
-2.08  & 1.29  &0.184      
&$4.26 \times 10^{-14}$
&0.957  &1.121  &0.036  &0.130  &0.053 \\
4.0  &4.0  &A &-1.64  &-0.247   &0.0152   &0.536   
&$5.83 \times 10^{-12}$ &
-2.09  & 1.58  &0.183      
&$3.94 \times 10^{-14}$
&0.933  &1.145  &0.037  &0.148  &0.060 \\
5.5  &0.0  &A &-2.35  &-0.668   &0.148    &1.96   
&$4.82 \times 10^{-14}$ &
-1.57  & 1.10  &0.198      
&$4.24 \times 10^{-14}$
&0.889  &1.076  &0.034  &0.110  &0.043 \\
5.5  &1.0  &A &-2.12  &-0.670   &0.0686    &1.35   
&$3.65 \times 10^{-13}$ &
-1.57  & 1.25  &0.197      
&$4.00 \times 10^{-14}$
&0.848  &1.078  &0.034  &0.115  &0.043 \\
5.5  &2.0  &A &-1.94  &-0.853   &0.0786   &0.921   
&$2.57 \times 10^{-13}$ &
-1.55  & 1.33  &0.195      
&$4.05 \times 10^{-14}$
&0.804  &1.095  &0.032  &0.118  &0.044 \\
5.5  &3.0  &A &-1.61  &-0.722   &0.0418   &0.720
&$7.58 \times 10^{-13}$ &
-1.59  & 2.12  &0.193   
&$3.20 \times 10^{-14}$
&0.768  &1.118  &0.033  &0.128  &0.049 \\
4.0  &0.0  &B &-2.62  &-0.0144  &0.0187   &5.74   
&$6.46 \times 10^{-12}$ &
-2.01  & 0.894  &0.198      
&$4.95 \times 10^{-14}$
&...  &...  &0.011  &0.042  &... \\
4.0  &1.0  &B &-2.52  &-0.0541  &0.0366   &6.65   
&$1.08 \times 10^{-12}$ &
-2.11  & 1.58  &0.197      
&$3.69 \times 10^{-14}$
&...  &...  &0.011  &0.043  &... \\
4.0  &2.0  &B &-2.36  &-0.0957   &0.0305   &6.44   
&$1.62 \times 10^{-12}$ &
-2.05  & 1.19  &0.197      
&$4.37 \times 10^{-14}$
&...  &...  &0.011  &0.042  &... \\
4.0  &3.0  &B &-2.09  &-0.193   &0.0199   &4.60   
&$4.21 \times 10^{-12}$ &
-2.10  & 1.64  &0.198      
&$3.63 \times 10^{-14}$
&...  &...  &0.011  &0.044  &... \\
4.0  &4.0  &B &-1.96  &-0.813   &0.0693   &3.48   
&$2.95 \times 10^{-13}$ &
-2.11  & 2.10  &0.198      
&$3.13 \times 10^{-14}$
&...  &...  &0.017  &0.056  &... \\
5.5  &0.0  &B &-2.80  &0.0356   &0.0203    &3.43   
&$2.74 \times 10^{-12}$ &
-1.09  & -0.370  &0.218      
&$1.17 \times 10^{-13}$
&...  &...  &0.017  &0.092  &... \\
5.5  &1.0  &B &-2.67  &0.0129   &0.0134   &3.44   
&$7.25 \times 10^{-12}$ &
-1.14  & -0.195  &0.216      
&$1.05 \times 10^{-13}$
&...  &...  &0.017  &0.088  &... \\
5.5  &2.0  &B &-2.45  &-0.00132  &0.0275   &5.14   
&$8.79 \times 10^{-13}$ &
-1.08  & -0.336  &0.216      
&$1.17 \times 10^{-13}$
&...  &...  &0.017  &0.085  &... \\
5.5  &3.0  &B &-1.90  &-0.0517   &0.0120   &7.28   
&$2.86 \times 10^{-12}$ &
-1.13  & -0.109  &0.211      
&$1.04 \times 10^{-13}$
&...  &...  &0.017  &0.082  &... \\
\enddata
\tablenotetext{a}{See equations (\ref{eqn:gradist}) and (\ref{eqn:sildist}).
In all cases, we take $a_{\rm c,s}=0.1 \micron$.}
\tablenotetext{b}{$R_V = A(V)/E_{B-V}$, ratio of visual extinction to 
reddening}
\tablenotetext{c}{C abundance in double log-normal very small grain
population (see equations \ref{eq:lognormal} and \ref{eq:B})}
\tablenotetext{d}{Total grain volumes in the carbonaceous
and silicate populations, normalized to their abundance/depletion-limited
values ($2.07 \times 10^{-27}$ and $2.98 \times
10^{-27} \cm^3 \, {\rm H}^{-1}$, respectively)}
\tablenotetext{e}{$\chi_1^2 = \sum_i (\ln A_{\rm obs} - \ln A_{\rm mod})^2
/\sigma_i^2$, for 100 points equally spaced in $\ln \lambda$}
\tablenotetext{f}{$\chi_2^2 = \sum_i (\ln A_{\rm obs} - \ln A_{\rm mod})^2$}
\tablenotetext{g}{$\chithree^2 = \chi_1^2 + 0.4 (\tilde{V}_{\rm g} -1)^{1.5} 
+ 0.4 (\tilde{V}_{\rm s} -1)^{1.5}$}
\end{deluxetable}

\begin{deluxetable}{ccccccccccccccc}
\tabletypesize{\scriptsize}
\rotate
\tablewidth{0pt}
\tablecaption{Grain Size Distribution Parameter 
Values for HD 210121\tablenotemark{a} 
\label{tab:HD210121}}
\tablehead{
\colhead{$10^5 b_{\rm C}$\tablenotemark{b}}&
\colhead{$\alpha_{\rm g}$}&
\colhead{$\beta_{\rm g}$}&
\colhead{$a_{\rm t,g}$}&
\colhead{$a_{\rm c,g}$}&
\colhead{$C_{\rm g}$}&
\colhead{$\alpha_{\rm s}$}&
\colhead{$\beta_{\rm s}$}&
\colhead{$a_{\rm t,s}$}&
\colhead{$C_{\rm s}$}&
\colhead{$\tilde{V}_{\rm g}$\tablenotemark{c}}&
\colhead{$\tilde{V}_{\rm s}$\tablenotemark{c}}&
\colhead{$\chi_1^2$\tablenotemark{d}}&
\colhead{$\chi_2^2$\tablenotemark{e}}&
\colhead{$\chithree^2$\tablenotemark{f}}
\\
\colhead{}&
\colhead{}&
\colhead{}&
\colhead{($\micron$)}&
\colhead{($\micron$)}&
\colhead{}&
\colhead{}&
\colhead{}&
\colhead{($\micron$)}&
\colhead{}&
\colhead{}&
\colhead{}&
\colhead{}&
\colhead{}&
\colhead{}
}
\startdata
0.0  &-2.22  &-0.0960  &0.00544  &0.651  &$1.71 \times 10^{-10}$ &
-1.96  &-5.23  &0.0999  &$2.32 \times 10^{-12}$
&0.752  &1.407  &0.071  &0.080  &0.175 \\
1.0  &-2.18  &-0.0818  &0.00551  &0.614  &$1.28 \times 10^{-10}$ &
-1.98  &-5.25  &0.105  &$1.99 \times 10^{-12}$
&0.745  &1.415  &0.070  &0.078  &0.177 \\
2.0  &-2.04  &-0.137   &0.00731  &0.566  &$5.37 \times 10^{-11}$ &
-1.96  &-6.05  &0.110  &$1.97 \times 10^{-12}$
&0.736  &1.423  &0.069  &0.077  &0.179 \\
3.0  &-1.87  &-0.190   &0.00911  &0.492  &$2.40 \times 10^{-11}$ &
-1.94  &-6.99  &0.112  &$2.09 \times 10^{-12}$
&0.726  &1.428  &0.072  &0.082  &0.184 \\
4.0  &-1.69  &-0.264   &0.0126  &0.449  &$8.60 \times 10^{-12}$ &
-1.90  &-9.22  &0.119  &$2.26 \times 10^{-12}$
&0.715  &1.442  &0.077  &0.088  &0.194 \\
\enddata
\tablenotetext{a}{See equations (\ref{eqn:gradist}) and (\ref{eqn:sildist}).
In all cases, we take $a_{\rm c,s}=0.1 \micron$.}
\tablenotetext{b}{C abundance in double log-normal very small grain
population (see equations \ref{eq:lognormal} and \ref{eq:B})}
\tablenotetext{c}{Total grain volumes in the carbonaceous
and silicate populations, normalized to their abundance/depletion-limited
values ($2.07 \times 10^{-27}$ and $2.98 \times
10^{-27} \cm^3 \, {\rm H}^{-1}$, respectively)}
\tablenotetext{d}{$\chi_1^2 = \sum_i (\ln A_{\rm obs} - \ln A_{\rm mod})^2
/\sigma_i^2$, for 100 points equally spaced in $\lambda^{-1}$}
\tablenotetext{e}{$\chi_2^2 = \sum_i (\ln A_{\rm obs} - \ln A_{\rm mod})^2$}
\tablenotetext{f}{$\chithree^2 = \chi_1^2 + 0.4 (\tilde{V}_{\rm g} -1)^{1.5} 
+ 0.4 (\tilde{V}_{\rm s} -1)^{1.5}$}
\end{deluxetable}

\begin{deluxetable}{cccccccccccccccc}
\tabletypesize{\scriptsize}
\rotate
\tablewidth{0pt}
\tablecaption{Size Distribution Parameter 
Values for the Magellanic Clouds\tablenotemark{a} 
\label{tab:grdistpars_MC}}
\tablehead{
\colhead{Environment}&
\colhead{$10^5 b_{\rm C}$\tablenotemark{b}}&
\colhead{$\alpha_{\rm g}$}&
\colhead{$\beta_{\rm g}$}&
\colhead{$a_{\rm t,g}$}&
\colhead{$a_{\rm c,g}$}&
\colhead{$C_{\rm g}$}&
\colhead{$\alpha_{\rm s}$}&
\colhead{$\beta_{\rm s}$}&
\colhead{$a_{\rm t,s}$}&
\colhead{$C_{\rm s}$}&
\colhead{$\tilde{V}_{\rm g}$\tablenotemark{d}}&
\colhead{$\tilde{V}_{\rm s}$\tablenotemark{d}}&
\colhead{$\chi_1^2$\tablenotemark{e}}&
\colhead{$\chi_2^2$\tablenotemark{f}}&
\colhead{$\chi^2$\tablenotemark{g}}
\\
\colhead{}&
\colhead{}&
\colhead{}&
\colhead{}&
\colhead{($\micron$)}&
\colhead{($\micron$)}&
\colhead{}&
\colhead{}&
\colhead{}&
\colhead{($\micron$)}&
\colhead{}&
\colhead{}&
\colhead{}&
\colhead{}&
\colhead{}&
\colhead{}
}
\startdata
LMC avg  &0.0  &-2.91  &0.895  &0.578  &1.21    &$7.12 \times 10^{-17}$ &
-2.45  &0.125  &0.191      
&$1.84 \times 10^{-14}$ &0.401  &0.675  &0.025  &0.069 &0.025 \\
LMC avg  &1.0  &-2.99  &2.46   &0.0980  &0.641    &$3.51 \times 10^{-15}$ &
-2.49  &0.345  &0.184   
&$1.78 \times 10^{-14}$ &0.330  &0.687  &0.018  &0.033 &0.018 \\
LMC avg  &2.0  &4.43    &0.0    &0.00322  &0.285  &$9.57 \times 10^{-24}$ &
-2.70  &2.18   &0.198   
&$7.29 \times 10^{-15}$ &0.279  &0.758  &0.016  &0.019 &0.016 \\
LMC 2    &0.0  &-2.94  &5.22  &0.373  &0.349    &$9.92 \times 10^{-17}$ &
-2.34  &-0.243  &0.184   
&$3.18 \times 10^{-14}$ &0.263  &0.753  &0.025  &0.043 &0.025 \\
LMC 2    &0.5  &-2.82  &9.01   &0.392  &0.269   &$6.20 \times 10^{-17}$ &
-2.36  &-0.113  &0.182   
&$3.03 \times 10^{-14}$ &0.252  &0.765  &0.022  &0.037 &0.022 \\
LMC 2    &1.0  &4.16   &0.0   &0.342  &0.0493  &$3.05 \times 10^{-15}$ &
-2.44  &0.254  &0.188   
&$2.24 \times 10^{-14}$ &0.206  &0.820  &0.012  &0.014 &0.012 \\
SMC bar  &0.0  &-2.79 &1.12 &0.0190 &0.522    &$8.36 \times 10^{-14}$ &
-2.26  &-3.46  &0.216   
&$3.16 \times 10^{-14}$ &0.254  &1.308  &0.017  &0.019 &0.027 \\
\enddata
\tablenotetext{a}{See equations (\ref{eqn:gradist}) and (\ref{eqn:sildist}).
In all cases, we take $a_{\rm c,s}=0.1 \micron$.}
\tablenotetext{b}{C abundance in double log-normal very small grain
population (see equations \ref{eq:lognormal} and \ref{eq:B})}
\tablenotetext{d}{Total grain volumes in the carbonaceous
and silicate populations, normalized to their abundance/depletion-limited
values (1.29, 1.86, 0.518, and 0.745 $\times 10^{-27} 
\cm^3 \, {\rm H}^{-1}$ for carbonaceous in LMC, silicate in LMC,
carbonaceous in SMC, and silicate in SMC, respectively)}
\tablenotetext{e}{$\chi_1^2 = \sum_i (\ln A_{\rm obs} - \ln A_{\rm mod})^2
/\sigma_i^2$, for 100 points equally spaced in $\lambda^{-1}$.}
\tablenotetext{f}{$\chi_2^2 = \sum_i (\ln A_{\rm obs} - \ln A_{\rm mod})^2$}
\tablenotetext{g}{$\chithree^2 = \chi_1^2 + 0.4 (\tilde{V}_{\rm g} -1)^{1.5} 
+ 0.4 (\tilde{V}_{\rm s} -1)^{1.5}$}
\end{deluxetable}

\end{document}